\documentclass[journal=nalefd,manuscript=letter,layout=traditional]{achemso}

\usepackage[utf8]{inputenc}

\usepackage{braket}
\usepackage{amsmath}
\usepackage{amssymb}
\usepackage{graphicx}
\usepackage{color}
\usepackage{physics}
\usepackage{siunitx}
\usepackage{xcolor}
\usepackage{soul}
\usepackage{gensymb}
\usepackage{comment}

\author{Sudipta Kundu}
\affiliation{Department of Materials Science and Engineering, Stanford University, Stanford, CA 94305, USA}
\author{Felipe H. da Jornada}
\email{jornada@stanford.edu}
\affiliation{Department of Materials Science and Engineering, Stanford University, Stanford, CA 94305, USA}
\affiliation{Stanford Institute for Materials and Energy Sciences, SLAC National Accelerator Laboratory, Menlo Park, CA 94025, USA}

\title{Exchange-mediated exciton splitting and linear dichroism in monolayer transition metal dichalcogenide induced by ferroelectric substrates}

\begin{document}
\begin{abstract}

Valley-polarized excitons in two-dimensional transition metal dichalcogenides (TMDs) offer a promising platform for quantum applications, yet the addressability and decoherence of these states remain fundamental challenges.
Here, by developing a first-principles electrostatic embedding approach and performing large-scale GW plus Bethe-Salpeter equation calculations, we reveal novel excitons that emerge in TMD monolayers when supported by a ferroelectric twisted bilayer hBN substrate.
We predict two competing low-energy excitons whose ordering depends on the dielectric environment: optically dark, charge-transfer excitons, and quasi-one-dimensional Wannier excitons with linear optical dichroism.
The spatial localization of Wannier excitons, together with intervalley exchange interactions in monolayer TMDs, splits valley-degenerate excitons by about 3~meV without external magnetic fields.
Our \textit{ab initio} calculations clarify the role of the interfacial twist angle and the spatial localization of fringe fields, establishing design rules for engineering long-lived two-level systems in TMD monolayers supported by ferroelectric substrates.
\end{abstract}

\maketitle

\section{Introduction}
Excitons in atomically thin transition metal dichalcogenides (TMDs) have garnered significant attention for more than a decade~\cite{qiu2013optical,chernikov2014exciton,wang2018colloquium}. Strong spin-orbit coupling and broken inversion symmetry enforce a valley-dependent optical selection rule in monolayer TMD whereby excitons in the $\textrm{K}$ and $\textrm{K}^{\prime}$ valleys couple to right- and left-circularly polarized light, respectively, making valley excitons attractive candidates for valleytronic and quantum-information applications. However, electron-hole exchange interactions and phonon-mediated intervalley scattering cause valley decoherence on picosecond timescales~\cite{yu2014valley,glazov2015spin,kioseoglou2016optical,lin2022phonon,jiang2021real}, motivating strategies to extend valley coherence.

Spatial confinement of excitons is one such strategy, as it reduces radiative recombination and can suppress exciton-phonon scattering, in some cases extending valley coherence times~\cite{wang2024quantum,durmucs2023prolonged}. Confinement of excitons has been demonstrated in TMD-on-TMD moir\'e superlattices~\cite{tran2019evidence,seyler2019signatures,kundu2023exciton,mak2022semiconductor,huang2022excitons,wilson2021excitons}, in monolayer TMDs subjected to spatially-varying gate potentials~\cite{thureja2022electrically,heithoff2023valley,hu2024quantum}, in TMDs deposited on patterned nano-pillars~\cite{wang2021highly}, in defect-bound exciton states~\cite{he2015single,chakraborty2015voltage,srivastava2015optically}, and in moir\'e patterns generated by anisotropic strain~\cite{bai2020excitons}. These approaches can break the underlying $C_{3v}$ symmetry of pristine TMDs and can yield linearly polarized emission as well as a splitting of the otherwise valley-degenerate bright excitons, typically in the range of 0.5--2~meV~\cite{thureja2022electrically,hu2024quantum,heithoff2023valley,wang2021highly}. Spatial confinement can also narrow the exciton linewidth down to $\sim$0.5~meV \cite{thureja2022electrically}, making these linearly polarized excitons individually addressable.

A promising platform for engineering exciton localization involves stacking a TMD monolayer on an hBN substrate displaying a small interfacial twist angle, the simplest realization being a twisted bilayer hBN (t-hBN). For twist angles $\lesssim 1\si\degree$, the t-hBN undergoes significant atomic reconstruction, leading to the emergence of large triangular AB and BA ferroelectric domains separated by sharp domain walls (DWs)~\cite{yasuda2021stacking,lv2022spatially,woods2021charge,zhao2021universal,zhou2022analyticaltheorynearfieldelectrostatic,kim2024electrostatic}. Such AB and BA domains give rise to sizable out-of-plane electric fields that imprint an alternating electrostatic potential on the supported TMD. Recent experiments on this geometry have observed Stark-shifted excitons spatially confined at the DWs~\cite{kim2024harnessing,gu2025quantum} with linearly polarized emission~\cite{gu2025quantum} instead of the traditional circular optical dichroism when excitons at the K and K' valleys are detuned through an external, out-of-plane magnetic field.

Despite the experimental success in demonstrating Stark-shifted excitons from such ferroelectric domains, there are significant fundamental questions relevant to this setup. There is disagreement on the experimental value of the electrostatic potential imprinted by the t-hBN, with reports ranging from 50--240~meV~\cite{zhao2021universal,kim2024electrostatic,li2017binary, yasuda2021stacking,woods2021charge} to as large as 400~meV~\cite{cho2024moire}. 
There is also no clear consensus on how the electrostatic potential depends on the twist angle or moiré length scale, with some studies reporting a strong dependence~\cite{kim2024electrostatic,cho2024moire}, while others observe no dependence at all~\cite{woods2021charge}.
Additionally, the fundamental microscopic origin of the exciton localization is not well understood, hindering the development of protocols to optimize exciton trapping. Notably, a first-principles understanding of how the t-hBN potential reshapes the excitonic structure of the proximitized TMD -- what sets the magnitude of the splittings, how spatial confinement of the Wannier exciton interplays with the underlying many-body effects, and what role the surrounding dielectric environment plays -- is still missing.

Here, we use large-scale first-principles GW plus Bethe-Salpeter equation (GW-BSE) calculations~\cite{rohlfing2000electron,deslippe2012berkeleygw} to predict the excitonic structure of a monolayer MoS$_2$ proximitized by t-hBN. We find two qualitatively distinct low-energy exciton families in the proximitized MoS$_2$: bright, quasi-one-dimensional (1D) Wannier excitons ($X^W$) localized within 2~nm of the DW regions, and dark, charge-transfer excitons ($X^{CT}$) in which the electron and hole reside at opposite sides of the alternating potential. 
The $X^W$ excitons display a significant energy splitting of $\sim3.1$~meV for states polarized longitudinally and transversely along the DW. We find this splitting originates from a unique cooperative effect between spatial confinement and many-body exchange interactions, the latter of which were previously predicted to affect the exciton dispersion at finite center-of-mass (COM) wavevectors~\cite{qiu2015nonanalyticity,glazov2015spin,yu2014dirac}, but which become finite at optical excitations due to the exciton COM localization.

Our calculations reveal that, compared to typical localization from lithographically defined gate setups, the large exciton splitting induced from t-hBN is dominated by the unusually narrow ($\sim$2~nm) DW, which makes the Stark confinement effective in a regime where the long-range exchange interaction is still relevant. Importantly, our calculations show that the position of the twisted hBN interface in thicker hBN substrates plays a key role, while the effect of the twist angle plays a marginal role below 1\si\degree. Instead, the twist angle primarily dictates the spatial decay of the electrostatic potential. Finally, to make these calculations tractable, we have also developed an \textit{ab initio} electrostatic embedding approach, wherein the t-hBN is integrated out by computing its effect on the TMD as a single proximity-induced effective potential at the ground-state density-functional theory (DFT) calculations, expanding the class of moir\'e-scale heterostructures that first-principles methods can address. Together, our results provide a microscopic basis for recent measurements~\cite{kim2024harnessing,cho2024moire,fraunie2023electron,gu2025quantum} and identify the dielectric environment, the t-hBN twist angle, and the substrate-TMD spacing as practical levers for engineering exciton localization and valley physics in two-dimensional materials.

\subsection{Proximity-induced potential in TMD monolayer}
We first review the physics of electrostatic potentials created by twisted hBN, and present our results from our multi-scale calculations based on parametrized force fields and first-principles calculations to deduce the effect that such a nearby twisted hBN imprints on the band structure of monolayer MoS$_2$.

\begin{figure}
    \centering
    \includegraphics[scale=0.25]{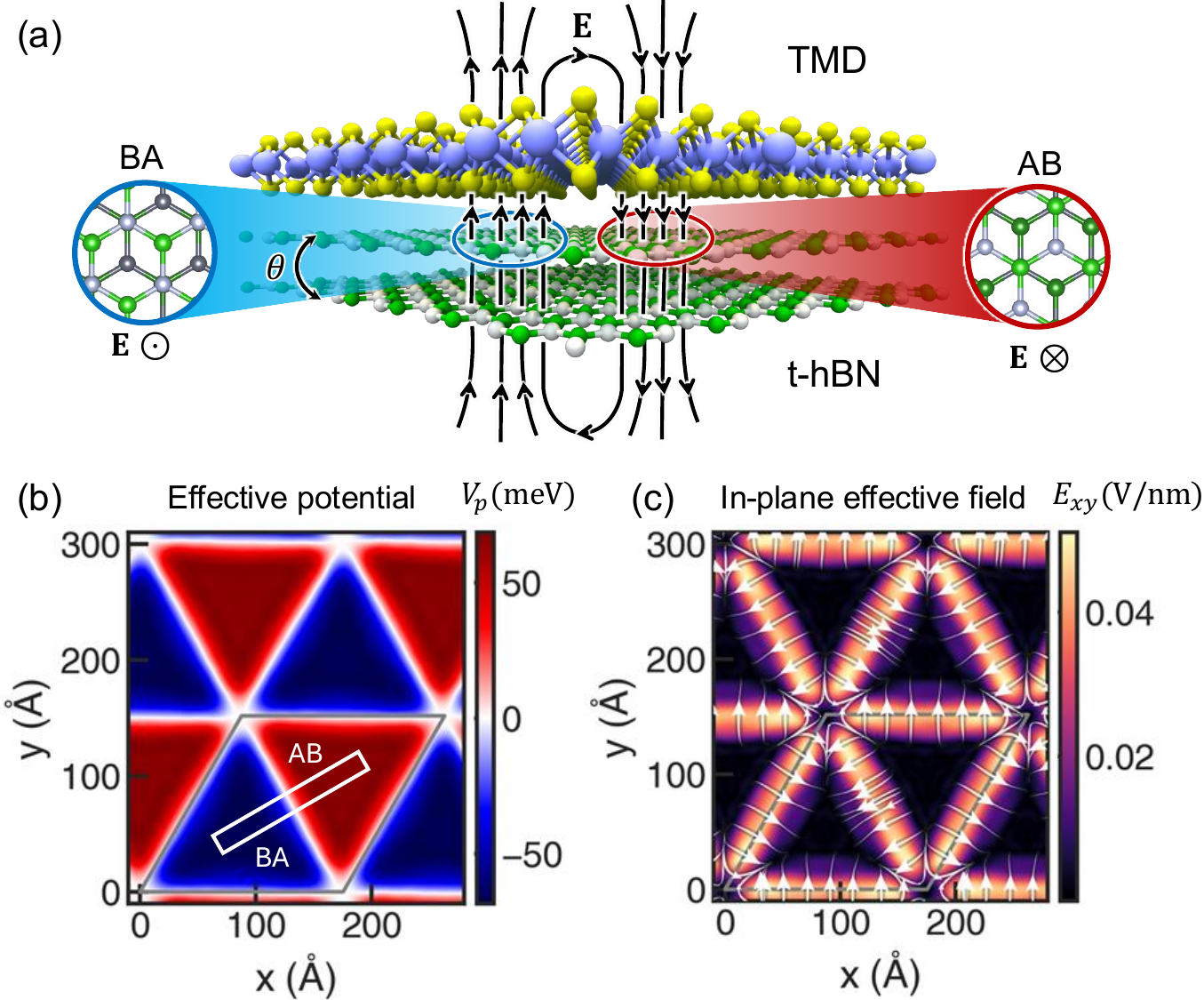}
    \caption{(a): Schematic of the device arrangement consisting of a t-hBN and a monolayer TMD, for example, MoS$_2$. The left and right circular insets show the two ferroelectric stackings of bilayer hBN: AB and BA with alternating dipoles, which give rise to electric fields along the alternating direction perpendicular to the hBN plane. The black lines with arrows penetrating through the heterostructure represent the spatially varying electric field originating from the t-hBN.
    (b) Polarization potential map $V_p$ due to the varying dipoles in a 0.8\si\degree t-hBN. The large red and blue triangles correspond to the AB and BA domains, while the thin white region separating them is the domain wall (DW).
    The white rectangle represents the one-dimensional strip across the DW. (c): The in-plane electric field in the plane at the twisted interface. 
    The arrows represent the direction, while the magnitude is represented by the color bar.
    }
    \label{fig:dft}
\end{figure}

A parallelly stacked bilayer hBN with a small, near-zero interlayer twist angle hosts large domains with alternating out-of-plane dipole polarization \cite{yasuda2021stacking}. The large domains consist of Bernal AB and BA stackings in which either a B or N atom from a reference top layer occludes the hollow site of the bottom layer, leading to oppositely aligned electric dipoles along the out-of-plane direction (Fig. \ref{fig:dft} (a) inset). Due to the Bernal stackings being the lowest energy stackings, they occupy the largest triangular-shaped regions in the t-hBN, and are separated by a DW (Fig.~\ref{fig:dft}(b)). The DW does not show any dipole along the out-of-plane direction \cite{bennett2023polar} owing to symmetry constraints. Another high-symmetry stacking, AA, possesses no dipole and will not be discussed here. 

The spatially varying dipole -- changing from a positive value we assign to the AB region to a negative value at the BA region -- gives rise to a spatially varying electrostatic potential. 
Using a multi-scale computational approach, we predict the effect of the t-hBN on the band structure of the TMD layer -- MoS$_2$ in our calculations. We first relax the large t-hBN structure with a classical force-field calculation as implemented in LAMMPS \cite{plimpton1995fast}. Subsequently, employing a series of DFT calculations \cite{kohn1996density} at various local stacking configurations of the t-hBN/MoS$_2$ heterostructure, we quantify the local change in the band structure of MoS$_2$. In particular, we ascribe the spatial variation of the local band edges of MoS$_2$ to an \emph{effective} potential $V_p$ experienced by the band edges in MoS$_2$. Because our \textit{ab initio} calculations explicitly include t-hBN and MoS$_2$, $V_p$ captures the small but finite electronic hybridization between MoS$_2$ and t-hBN (see SI).

Figure \ref{fig:dft}(b) depicts the distribution of $V_p$ for a 0.8\si\degree{} t-hBN/MoS$_2$ heterostructure.
The large red and blue triangular domains ($\sim$ 100 \AA{}) originate from the stable AB and BA stackings of the nearby t-hBN structure. Our \textit{ab initio} calculations reveal that $V_p$ has a potential width of 148 meV, which corresponds to the maximum band offset difference of MoS$_2$ in the presence of the supporting t-hBN. We find a potential modulation of 187 meV in only t-hBN, which is comparable to the previously reported potential modulation width of $\sim$200 meV \cite{zhao2021universal,kim2024electrostatic,li2017binary}, highlighting the importance of accounting for the screening from MoS$_2$. We note here that some experimental studies reported a slightly larger potential width of 225-240 meV \cite{yasuda2021stacking,woods2021charge}, which may arise from stronger interlayer interactions in structures with thicker hBN substrates on both sides of the interface. In another study \cite{cho2024moire}, a surprisingly large potential width of 400 meV was reported following annealing of t-hBN. Our calculations indicate that a value of this magnitude cannot be accounted for by interlayer electronic hybridization alone, and likely reflects additional contributions from structural reconstruction or the presence of free carriers. Another important observation is that $V_p$ is nearly zero for a region $\sim$ 20\AA{} wide, which closely corresponds to the width of the DW in t-hBN.

The spatial variation of the potential induced by the spatially varying dipoles in t-hBN further gives rise to an electric field in MoS$_2$. Similarly to our previous analysis, we can define an \emph{effective} in-plane field from the gradient of the effective potential experienced by MoS$_2$, $\mathbf{E}_\mathrm{eff}=\nabla_{xy} V_p$. For small twist angles, $V_p$ quickly saturates to a constant value inside the large AB and BA domains; hence, $\mathbf{E}_\mathrm{eff}$ is non-zero only at the DW region (Fig. \ref{fig:dft}(c)), with a maximum value of 50 mV/nm.

A critical issue now is in identifying the twist angle of the bilayer hBN that warrants investigation for novel excitonic states.
In an experimental condition, the twist angle of the t-hBN can be made almost arbitrarily small, $\lesssim$ 0.1\si\degree{}, resulting in large domain sizes $\gtrsim$ 100 nm. This represents a well-defined limit wherein one expects interesting physics emerging from electronic states across the different AB and BA domains, which is precisely at the DW. Accordingly, we focus our calculations on this small twist-angle limit, for which the critical component is the variation of $V_p$ along the AB-DW-BA local domains (white rectangle in Fig. \ref{fig:dft}(b)). Consequently, this limit can be well-captured by a quasi-1D calculation, wherein one approximates the potential $V_p$ as being constant along one direction and displaying the alternating profile from the AB-DW-BA path.

To efficiently perform such calculations with a standard plane-wave basis, we apply the electrostatic embedding scheme outlined as follows: rather than including the t-hBN atoms explicitly, we integrate out the substrate by adding the \textit{ab initio}-deduced effective potential $V_p$ directly into a DFT calculation of the proximitized MoS$_2$ monolayer. The resulting DFT plus GW-BSE calculations are performed on a large, 300~\AA{}-long quasi-1D supercell that minimizes interactions between excitons localized at neighboring DWs and at their periodic replicas. Direct GW-BSE calculations on a fully atomistic TMD/t-hBN moir\'e supercell -- which would contain thousands of atoms -- remain prohibitive even on leadership-class facilities. On the other hand, the electrostatic embedding approach combined with the quasi-1D geometry reduces the computational cost by orders of magnitude while preserving the proximity-induced screening and self-energy effects that determine the excitonic spectrum, extending first-principles many-body calculations to a class of moir\'e-scale heterostructures otherwise out of reach.

\subsection{Electronic structure}
\begin{figure}[h]
    \centering
    \includegraphics[scale=0.28]{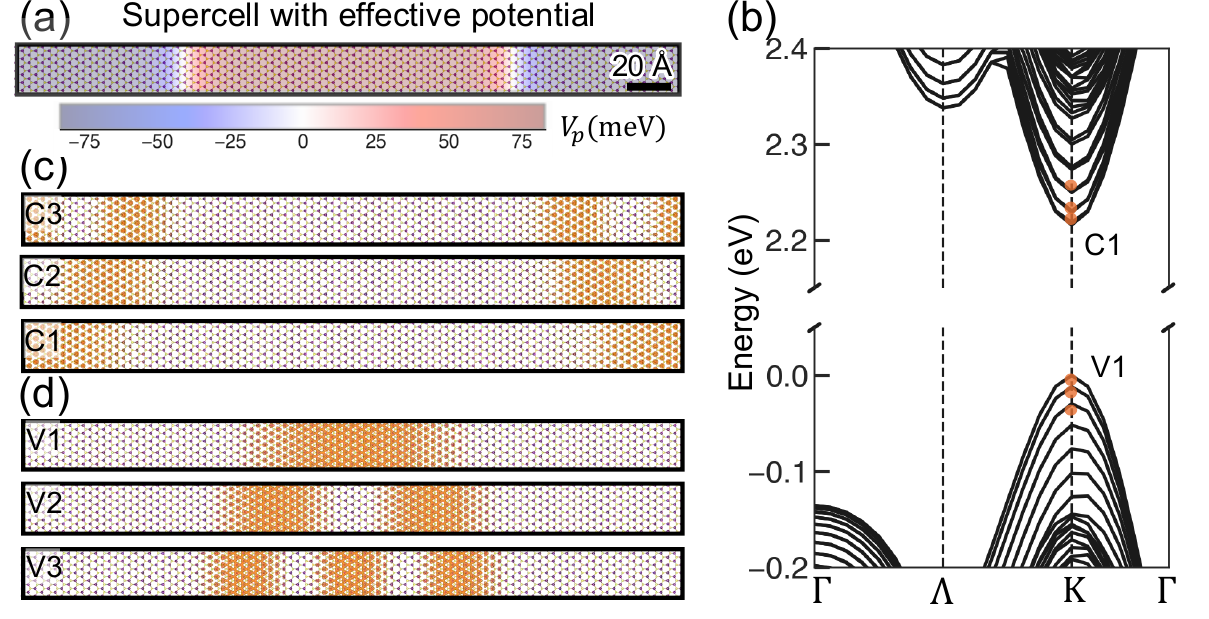}
    \caption{(a): Schematic of the 1D approximation for $V_p$. The underlying color profile represents the varying $V_p$ applied to the 1D supercell of MoS$_2$. (b): Band structure of the MoS$_2$ supercell including the effect of $V_p$ (c) and (d): $\lvert\psi_{K}(\textbf{r})\rvert^2$ of the first three non-degenerate bands at the valence and conduction band edges, respectively. The isosurface is plotted up to 0.0001 e/Bohr$^3$. }
    \label{fig:bands}
\end{figure}

We compute the electronic structure of the heterostructure using DFT within the Quantum Espresso package \cite{giannozzi2009quantum} with our electrostatic embedding approach.
We add the effective external potential $V_p$ when self-consistently solving the Kohn-Sham equations, properly rescaling $V_p$ to account for the spurious finite dipole induced in a supercell approach (see SI).
Furthermore, when defining a suitable unit cell that encompasses our quasi-1D MoS$_2$ geometry along with the effective potential $V_p$, we include a mirrored copy of $V_p$ to satisfy periodic boundary conditions (Fig.~\ref{fig:bands}(a))
While we consider here MoS$_2$ for our calculations, our main conclusions apply to all semiconducting TMD monolayers with similar band structures.

Our DFT plus GW calculations on this quasi-1D system with the effective potential $V_p$ reveal a number of qualitatively expected features. In the presence of $V_p$, the quasiparticle bandgap of TMD within the hBN-encapsulated dielectric environment decreases
from 2.36 eV to 2.21 eV, since valence and conduction states are now spatially separated and localized according to the regions where monolayer MoS$_2$ experiences either a positive or negative $V_p$, respectively.
The charge densities corresponding to the first three conduction band and valence band wavefunctions at the $\textrm{K}$ point ($\lvert\psi_{\textrm{K}}(\textbf{r})\rvert^2$) of the supercell Brillouin zone are shown in Fig. \ref{fig:bands}(c) and (d) respectively. The valence band maximum localizes at the center of the supercell, corresponding to the potential hill, and the conduction band edge localizes at the edges of the supercell (potential well). We note that the higher-energy holes and electrons display a nodal profile and spatial dependence that are qualitatively between those resulting from a 1D infinite-well and harmonic oscillator potentials. This highlights the importance of carefully capturing the effective potential $V_p$, including relaxation effects from the underlying t-hBN substrate.

\subsection{Excitons and absorption spectrum}
We investigate the excitons in the MoS$_2$ supercell including the effect of $V_p$ by solving the Bethe-Salpeter equation \cite{rohlfing2000electron} using the BerkeleyGW package \cite{deslippe2012berkeleygw},
\begin{equation}
\begin{split}
    \Delta_{cv\textbf{k}}A^S_{vc\textbf{k}} + \sum_{v^{\prime}c^{\prime}\textbf{k}^{\prime}}(\langle vck\lvert K^d\lvert v^{\prime}c^{\prime}\textbf{k}^{\prime}\rangle &+ \langle vck\lvert K^x\lvert v^{\prime}c^{\prime}\textbf{k}^{\prime}\rangle) \\ 
    &= \Omega^SA^S_{vc\textbf{k}}
\end{split}
\end{equation}
where $\Delta_{cv\textbf{k}}=E_{c\textbf{k}}-E_{v\textbf{k}}$ is the quasi-particle energy difference of the 
constituent electron ($c$) and hole ($v$) at k-point $\textbf{k}$, $K^d$, $K^x$ denote the direct and exchange interactions of the electron-hole pair ($\lvert vc\textbf{k}\rangle$) respectively and $\Omega^S$ is the energy of the exciton $S$ whereas $A^S_{vc\textbf{k}}$ are exciton expansion coefficients in the electron-hole basis. We employ a mixed stochastic-deterministic approach to speed up the calculation of the dielectric screening of the MoS$_2$ supercell \cite{altman2024mixed}. We include the additional screening from hBN or any other embedding dielectric media using a modified Keldysh form \cite{keldysh2024coulomb,da2020universal} (see SI). 

\begin{figure}[]
    \centering
    \includegraphics[scale=0.3]{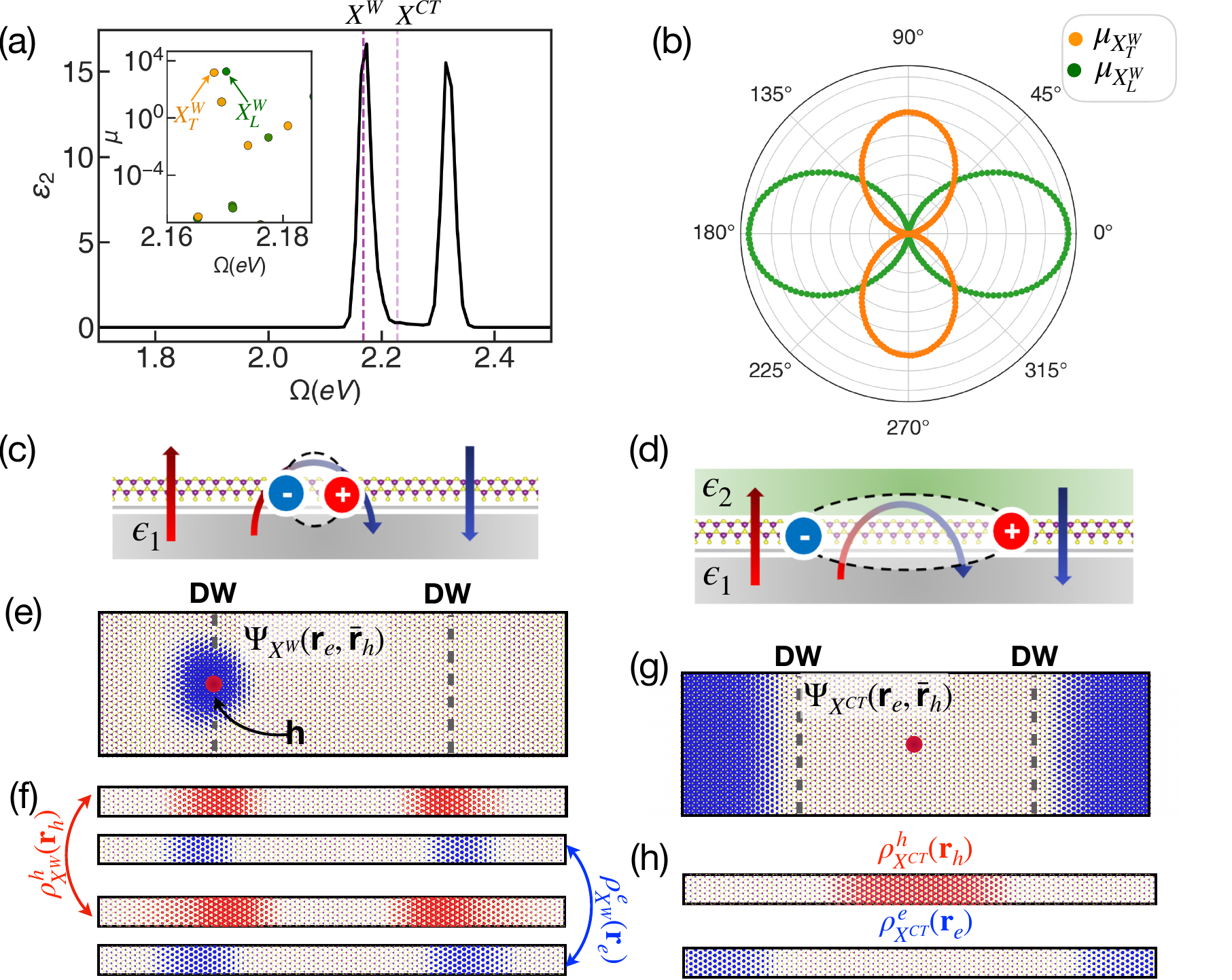}
    \caption{(a): 
    Absorption spectrum of MoS$_2$/t-hBN encapsulated with hBN.
    The vertical dashed lines indicate the positions of the lowest energy bright $X^W$ and the lowest energy $X^{CT}$, respectively. The inset focuses on the low energy bright $X^W$'s, highlighting the two types: transverse ($X^W_T$) and longitudinal ($X^W_L$) excitons by orange and green dots, respectively. (b): Polar plot of the oscillator strengths corresponding to the excitons shown in the inset. 0\si\degree direction corresponds to the direction of $V_p$. (c) and (d): Schematic of the electric field and probable excitons in the MoS$_2$/t-hBN  heterostructure stack when deposited on a bulk hBN substrate and when encapsulated by dielectric materials on both sides, respectively. $\epsilon_1$ refers to the dielectric medium of bulk hBN. $\epsilon_2$ can be the same as $\epsilon_1$ or a different dielectric. (e): Distribution of the exciton wavefunction ($\lvert\Psi(\textbf{r}_e,\Bar{\textbf{r}}_h)\rvert^2$) of $X^W$ when the hole is fixed at the DW. The electron cloud is shown in blue and the red dot represents the hole. (f): The distribution of the hole  (electron)-density while integrating over the electron (hole): $\rho^h_{X^W}(\textbf{r}_h)$ ($\rho^e_{X^W}(\textbf{r}_e)$) is shown in red (blue) for $X^W_T$ and $X^W_L$ respectively. (g) Same as (e) for $X^{CT}$ when the hole is fixed at the center of the supercell. (h): Hole and electron-density of $X^{CT}$ as defined for (f).}
    \label{fig:absp}
\end{figure}

Given the experimental advantage of encapsulating TMDs to reduce exciton linewidths, we investigate a t-hBN/MoS$_2$ heterostructure encapsulated by bulk hBN.
The absorption spectrum calculated with first-principles calculations shows the usual $A$ and $B$ peaks with an energy separation of 142 meV (Fig. \ref{fig:absp}(a)). Additionally, there are many excitons with varying oscillator strengths. We focus on the excitons corresponding to the bright $A$ peak (marked by a darker purple dashed line in Fig. \ref{fig:absp}(a)). These excitons are tightly bound Wannier excitons ($X^W$) and show a fine structure. While the lowest-energy spin-forbidden dark $X^W$'s are energetically four-fold degenerate, the bright $X^W$'s are not similar to those in a pristine MoS$_2$ \cite{qiu2015nonanalyticity}, and split into two distinct types of excitons: $X^W_T$ and $X^W_L$ (Fig. \ref{fig:absp}(a) inset). They couple to linearly polarized light instead of the usual circular polarization observed in pristine TMDs. The lower energy $X^W_T$ couples to photons with the polarization vector \emph{transverse} to the confining direction imprinted by $V_p$ while the higher energy $X^W_L$, to the \emph{longitudinal} direction (Fig. \ref{fig:absp}(b)). Both $X^W_T$ and $X^W_L$ are localized at the DW regions with different degrees of localization. The distribution of the electron density (blue) of the lowest-energy bright exciton ($\lvert\Psi_{X^W}(\textbf{r}_e,\bar{\textbf{r}}_h)\rvert^2$), when the hole (red dot) is fixed ($\bar{\textbf{r}}_h$) at the middle of DW, is shown in Fig. \ref{fig:absp}(e), which clearly reveals a bound Wannier exciton. This bound exciton with a binding energy of 196 meV further shows quasi-one-dimensional confinement at the DW regions.
The quasi-one-dimensional localization is further illustrated by plotting the hole (electron) density while integrating over the other particle: $\rho^h=\int \lvert\Psi_{X^W}(\textbf{r}_e,\textbf{r}_h)\rvert^2 \textbf{dr}_e$ ($\rho^e=\int \lvert\Psi_{X^W}(\textbf{r}_e,\textbf{r}_h)\rvert^2 \textbf{dr}_h$) (Fig.~\ref{fig:absp}(f)). Both $\rho^h$ and $\rho^e$ of $X^W_T$ and $X^W_L$ show localization at the DW region along the confinement direction while being continuous along the direction perpendicular to it. $X^W_T$'s are lower in energy, consequently more bound and have a smaller radius (15.12 \AA{}) compared to $X^W_L$'s (16.8 \AA{}). 

At higher energy, we observe a qualitatively different charge-transfer exciton $X^{CT}$ (lighter purple dashed line in Fig. \ref{fig:absp}(a)) in which the electron shows a large amplitude at the potential well region while the hole resides at the potential hill (Fig. \ref{fig:absp} (g)). Due to the spatial separation of the electron and hole (Fig. \ref{fig:absp} (h)), the oscillator strength of $X^{CT}$ is three orders of magnitude smaller than that of $X^W$.

The internal structures of these two types of excitons are different, and therefore, they respond differently to the external screening environment. The strongly bound $X^W$ possesses a large electron-hole spatial overlap and is more sensitive to the dielectric environment than the weakly bound $X^{CT}$, which display an energy that decreases approximately linearly with the bandwidth of the effective potential. 
To investigate the sensitivity of such states with dielectric environments, we perform another calculation wherein we considered a TMD stacked on bulk hBN, wherein the twisted hBN interface is the one closest to the TMD, but now without considering any additional encapsulating hBN on the other side of the TMD. 
 This reduced external screening in the system increases the electron-hole interaction strength. As a result, the binding energy of $X^W$ increases to 255 meV while the $X^{CT}$ energy changes only by 7 meV, significantly enlarging the energy difference between the lowest energy  $X^W$ and  $X^{CT}$ ($\Delta E^{W-CT}$) from 60 meV to 127 meV. 
We also consider increasing the dielectric constant of the encapsulating material on one side ($\epsilon=15$, corresponding to another TMD), which reduces the binding energy of the $X^W$, and therefore $\Delta E^{W-CT}$ to 22 meV. Further increasing the dielectric by gating or combination of gating with a higher dielectric material, it is possible to tune between the two types of excitons. 

The dielectric tunability of $\Delta E^{W-CT}$ also suggests a possible dynamical use of this system. Because $X^{CT}$ has its electron and hole spatially separated across the t-hBN potential, its oscillator strength is roughly three orders of magnitude smaller than that of $X^W$, and its radiative lifetime correspondingly longer. One could in principle optically populate $X^W$ in a dielectric environment where it is the lowest-energy state, and then dynamically modulate the surrounding screening -- for example through a gate-tunable dielectric layer -- to reorder $X^W$ and $X^{CT}$, transferring the optical excitation into the long-lived dark $X^{CT}$ as an optical memory. The proximity field strength, the substrate spacing, and the encapsulating dielectric provide tunable knobs over both the energy ordering and the spatial overlap of $X^W$ and $X^{CT}$, opening a route to engineering such on-demand bright-to-dark exciton conversion in TMD heterostructures.

Now, we discuss the origin of the fine structure of the $X^W$'s (Fig. \ref{fig:absp}(a)) and their localizations (Fig. \ref{fig:absp}(e)). The DWs, where $X^W$'s are confined, correspond to the region with the largest in-plane gradient of $V_p$, which we refer to as the effective in-plane electric field $E_{xy}$. Since the lowest-energy A excitons in MoS$_2$ have no finite electric dipole in their COM degrees of freedom, they can be localized through an effective DC Stark potential that is a second-order effect in $E_{xy}$ ($V_{\mathrm{eff}} \sim \lvert E_{xy}\rvert^2$).  
In a simpler picture, one can rationalize the optically active, zero wavevector $X^W$ confined at the DWs as a linear combination of unconfined excitons in a pristine MoS$_2$ monolayer with various wavevectors $\textbf{Q}$'s. The DC Stark potential ($V_{\mathrm{eff}}$)
mixes in excitons with wavevectors $\textbf{Q}$'s separated by a reciprocal lattice vector of our large 1D supercell, leading to the spatial confinement.

At $\textbf{Q}=0$ in pristine MoS$_2$, the lowest energy bright excitons are doubly degenerate and confined in each of the $\textrm{K}$ ($\lvert \Psi_{\textrm{K}}\rangle$) and $\textrm{K}^\prime$ ($\lvert \Psi_{\textrm{K}^\prime}\rangle$) valleys. Exchange interaction ($K^x$) is proportional to $\textbf{Q}$. At finite $\textbf{Q}$, mediated by the exchange interaction, the valley excitons can form in-phase and out-of-phase linear combinations, which we refer to as valley-bonding ($\lvert \Psi_{+}(\textbf{Q})\rangle \sim (\lvert \Psi_{\textrm{K}}^{\textbf{Q}}\rangle+\lvert \Psi_{\textrm{K}^\prime}^{\textbf{Q}}\rangle)$) and valley-antibonding ($\lvert \Psi_{-}(\textbf{Q})\rangle \sim (\lvert \Psi_{\textrm{K}}^{\textbf{Q}}\rangle-\lvert \Psi_{\textrm{K}^\prime}^{\textbf{Q}}\rangle)$) excitons respectively \cite{qiu2015nonanalyticity,glazov2015spin,yu2014dirac}. Valley-bonding excitons display a finite valley exchange interactions and a polarization that is longitudinally polarized with $\mathbf{Q}$, and, hence, acquire a linear dispersion ($E_{+}(\textbf{Q})$) with \textbf{Q} for small \textbf{Q}'s. On the other hand, the valley-antibonding are transverse excitons and only retain their parabolic dispersion ($E_{-}(\textbf{Q})$).  
The energy difference $\Delta E_{\pm}$ between the two exciton bands is driven by the $\textbf{Q}$ dependent exchange interaction. 

We can now rationalize the aforementioned splitting in a simple reciprocal-space picture. One can approximately express an exciton that is spatially localized in its COM coordinate by some length $\ell$ as a linear combination of unconfined excitons that are delocalized in momentum space by $\sim1/\ell$. Owing to the different band dispersion for the transverse ($\lvert \Psi_{-}(\textbf{Q})\rangle$) and longitudinal ($\lvert \Psi_{+}(\textbf{Q})\rangle$) exciton due to many-body exchange interactions, there is a higher kinetic energy cost for spatially localizing $X^W_{T}$ versus $X^W_{L}$, which gives rise to the energy split $\Delta E_{LT}$ between them even. In our hBN-encapsulated calculation involving a 0.8\si\degree t-hBN/MoS$_2$ heterostructure, $\Delta E_{LT}$ is 2.1 meV.
The exchange interaction further influences the localization and absorption characteristics of $X^W_T$ and $X^W_L$ (Fig. \ref{fig:absp}). Following the linear polarization behavior of $\lvert \Psi_{-}\rangle$ and $\lvert \Psi_{+}\rangle$ at finite $\textbf{Q}$, $X^W_{T}$ and $X^W_{L}$ also show linear polarization along directions orthogonal to each other. In addition, $X^W_L$ derives from the more dispersive linear-parabolic $\Psi_+(\textbf{Q})$ branch, and hence for a given $V_p$, $X^W_L$ is more 
delocalized in real space compared to $X^W_T$. This also leads to larger oscillator strength for $X^W_L$ as is evident from Fig. \ref{fig:absp}(b).

\begin{figure}[]
    \centering
    \includegraphics[scale=0.36]{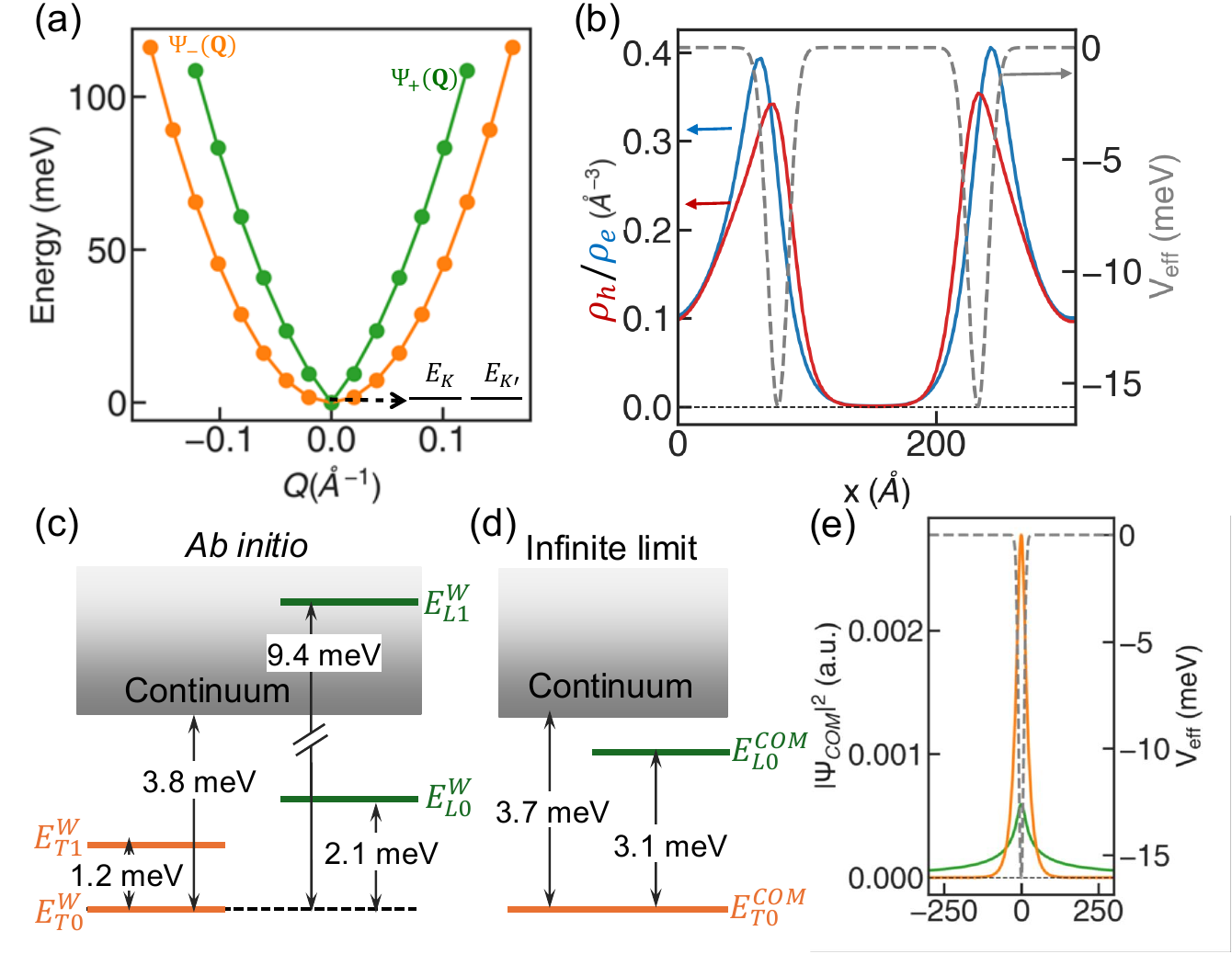}
    \caption{(a): Band dispersion of the valley antibonding (orange) and bonding (green) excitons in hBN encapsulated pristine TMDs. (b): Effective COM potential (grey) and the electron (blue) and hole (red) components of the \textit{ab initio} $X^W_T$ wavefunctions for hBN encapsulated heterostructure, averaged over two excitons localized at either DWs. (c) The \textit{ab initio} energies corresponding to the first two bright excitons of both transverse and longitudinal series of hBN encapsulation are shown, setting the continuum at zero.  (d) Exciton energies from the COM Hamiltonian model at the infinite limit. A schematic of the effective potential and the exciton wavefunctions in COM coordinates is shown.  }
    \label{fig:model}
\end{figure}

The qualitative picture above -- finite-$\textbf{Q}$ valley-bonding/antibonding excitons mixed by the supercell's Stark potential into linearly polarized confined states at the DW -- accounts for both the localization and the LT splitting found in our \textit{ab initio} calculations. To put it on a quantitative footing and to extrapolate to twist angles smaller than the 0.8\si\degree\ case directly accessible by GW-BSE, we further construct a minimal model Hamiltonian in the exciton COM coordinate and fit it to our \textit{ab initio} energies. We take the Hamiltonian to have the form $H_{\mathrm{COM}}(\textbf{Q})=E_{\pm}(\textbf{Q}) + V_{\textrm{eff}}(\textbf{Q})$.
Due to the expected Stark-like form, we constrain $V_{\mathrm{eff}}(\textbf{R})$ to be proportional to the \textit{ab initio} effective in-plane field, $V_{\mathrm{eff}}(\textbf{R}) = -\frac{1}{2}\alpha \lvert E_{xy}(\textbf{R})\rvert^2 $ with \textbf{R} denoting spatial coordinate of the supercell in real space and $\alpha$ is the polarization coefficient of the excitons. 
In order to obtain a quantitative understanding of the $\Delta E_{LT}$, we solve for the low-energy excitons in the COM frame. 
Fig. \ref{fig:model}(b) shows the $V_{\textrm{eff}}$ for hBN encapsulated heterostructure and $\rho_h$, $\rho_e$ of $X^W_{T}$, averaged over two excitons localized at both DW's, are overlaid on top of the potential. Fig. \ref{fig:model}(c) depicts the \textit{ab initio} energy levels for the two lowest $X^W_{T}$ and $X^W_{L}$'s with respect to the continuum 
for hBN encapsulation. The continuum energy is set by the lowest energy bright exciton in pristine TMD in the corresponding dielectric environment. To extract the effective depth of $V_{\mathrm{eff}}$, we fit for the bound exciton energies corresponding to $X^W_{T0}$, $X^W_{T1}$ and $X^W_{L0}$, obtaining a large $V_{\mathrm{eff}}$ of 16 meV. The exciton polarizability $\alpha$ is estimated to be 11.7 eV\,nm$^2$/V$^2$, which is larger to that reported in pristine TMD under a uniform electric field \cite{cavalcante2018stark}. This enhanced polarizability can originate from the confined exciton states at the domain wall, which are formed through the mixing of finite-\textbf{Q} pristine excitons by the localized electric field in the DW regions.

Employing our fitted $H_{COM}$ to a very small twist angle corresponding to $\sim 0.1\si\degree$, we find the $\Delta E_{LT}$ to be 3.1 meV. Here, we note that our calculations reveal that 
the width of the exciton confining potential well to be rather insensitive to the twist angle for $\theta \lesssim 1\si\degree$. 
This exchange-induced energy split between longitudinal and transverse polarized excitons (L-T splitting) is larger than what was observed earlier in gate-defined, electrostatically confined excitons \cite{thureja2022electrically,wang2021highly}. We find that our larger predicted L-T splitting does not originate from a larger potential; rather, it stems from the giant confinement of the Wannier excitons caused by the small size of the DW region, which can Stark confine excitons in a small region. 
This large L-T splitting allows one to address the excitons individually with a transverse or longitudinally polarized light. 
In addition, such a large detuning of valley excitons can help circumvent phonon- and exchange-mediated valley scattering in TMDs, suppressing an important exciton dephasing mechanism. 

\subsection{Discussion and conclusion} 

\begin{figure}
    \centering
    \includegraphics[scale=0.27]{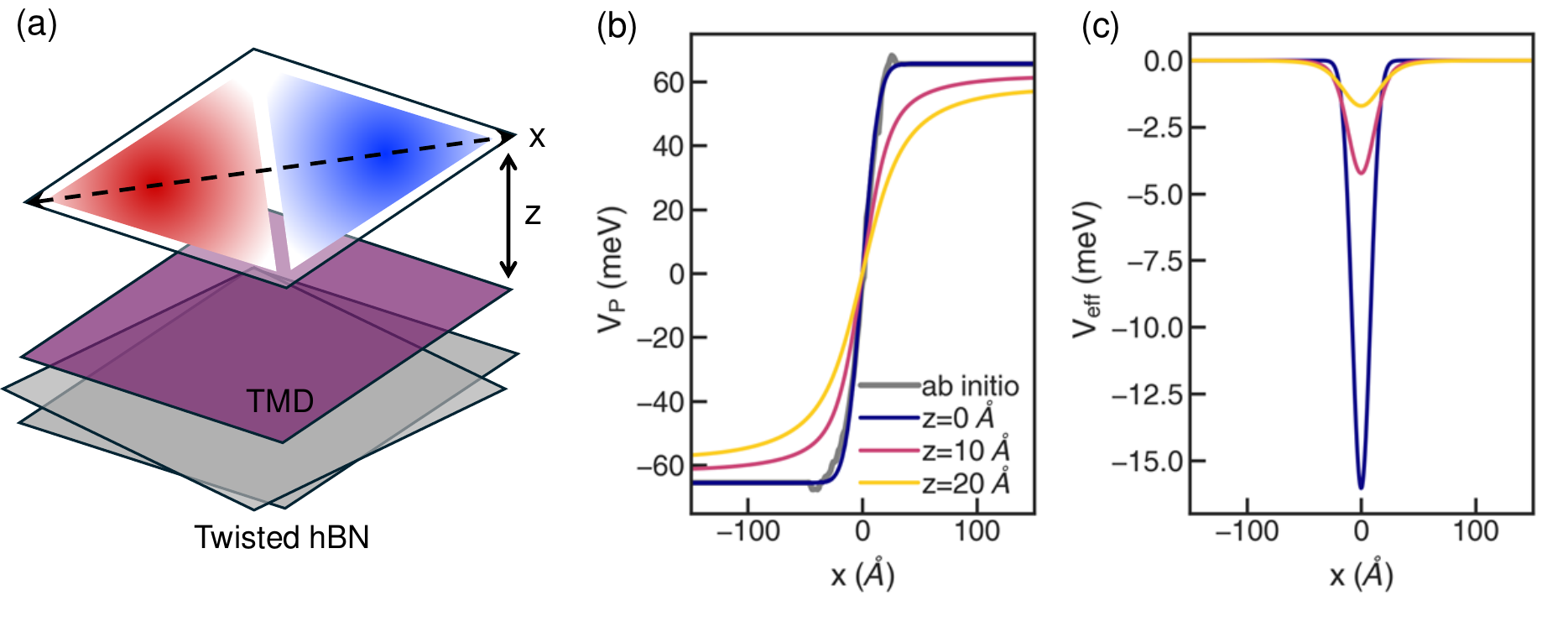}
    \caption{(a) Schematic of the device setup depicting the axes orientation. (b) and (c): Variation of $V_p$ and $V_{\textrm{eff}}$ with increasing $z$ from the equilibrium position of TMD on t-hBN.}
    \label{fig:dist}
\end{figure}

From our large-scale \textit{ab initio} calculations, it is evident that to achieve a large $\Delta E_{LT}$, one needs a strong
electric field in the TMD layer at a small length scale ($\sim$ 2 nm). Since the split is quadratically proportional to the electric field at the DW region, the distance between the twisted interface and the TMD layer becomes critical. The effective potential $V_p$ at any (\textbf{R},$z$) point (Fig. \ref{fig:dist}(a)), where $R$ is the in-plane position in the moir\'e supercell and $z$ is in out-of-plane direction, can be described by \cite{zhao2021universal}: 
\begin{align}
    V_p(\textbf{R},z) = \textrm{sgn}(z)\sum_{\textbf{G}}V_{\textbf{G}}e^{i\textbf{G}.\textbf{R}}e^{-G|z|}
\end{align}
\textbf{G} are the reciprocal lattice vector of the moir\'e supercell. Fig. \ref{fig:dist}(b) shows the variation of $V_p$ along the diagonal direction of the moire supercell near the DW for various $z$ values for $\theta=0.3 \si\degree$. The variation of $V_p$ across DW becomes smaller as the distance from the twisted interface increases. Consequently, the DC Stark potential $V_{\textrm{eff}}$ shows a rapid decay with increasing distance for a given $\theta$ (Fig. \ref{fig:dist}(c)). 
However, the in-plane electric field at the twisted interface is not sensitive to $\theta$ for $\theta \lesssim 1\si\degree$ since the DW does not change significantly with $\theta$ for such $\theta$'s. 
A strong and spatially confined electric field is desirable for a larger $\Delta E_{LT}$ and therefore, the split is best observed when the TMD is stacked on top of a small-angle twisted hBN without any spacer layer in between them. It is worth noting here that although the potential magnitude and its derivative at the interface do not depend on $\theta$ for small $\theta$'s, the out-of-plane decay rate is indeed affected by it. As $\theta$ increases and the domain size decreases, the electrostatic potential decays faster along the out-of-plane direction, which explains the domain size dependence of the measured potential in structures with thick hBN layers above the interface \cite{kim2024electrostatic}.

We discuss two different setups of experimental protocols for measuring the optical signature. First, one can use a setup of a monolayer TMD on top of t-hBN. In this case, it is crucial to have small twist angles and, correspondingly, large AB and BA domains, such that a single, diffraction-limited laser spot illuminates a single domain instead of several domains with alternating polarization directions unless one utilizes polarization-resolved near-field measurements \cite{zhou2024probing}. Such experimental setups have been explored recently, which report observations of Stark shift of confined excitons at the DW regions \cite{kim2024harnessing,gu2025quantum} and their linear polarization parallel to DW \cite{gu2025quantum}, consistent with our findings. We note that the absence of the signature of linear polarization in some cases might result from the averaging across multiple DW regions. Signatures of exciton-polaron have also been reported for large AB/BA domains, indicating charge migration in TMD from one domain to another and the formation of multi-particle ($> 2$) exciton complexes in the presence of extra charge \cite{fraunie2023electron,cho2024moire}, capturing which is beyond the scope of this study. 

Another possible setup is to engineer a 1D moir\'e by uniaxially straining a bilayer hBN \cite{han2024highly}. The advantage of this setup is that it would yield strictly parallel domains with well-defined polarization directions and would remove the requirement of the illumination spot to be larger than the domain size in order to observe linear polarization.


In conclusion, we develop an \textit{ab initio} electrostatic embedding scheme, perform large-scale GW-BSE calculations, and map our results to minimal effective Hamiltonians to understand the underlying excitonic states of MoS$_2$ supported by a ferroelectric t-hBN substrate. We find a lowest-energy quasi-1D Wannier exciton localized within just $\sim2$~nm -- a significant spatial confinement possible due to the narrow spatial features of the t-hBN DWs compared to the extent of lithographically defined fringe fields. Such strong spatial localization, in conjunction with exchange interaction, leads to the lifting of the exciton degeneracy by 3.1 meV, and the emergence of two optically addressable excitons with linear optical polarization.
The L-T splitting, much larger compared to the typical linewidth of confined excitons ($\sim$ 0.5 meV) in TMDs\cite{thureja2022electrically}, provides an accessible two-level system, suitable for application in quantum information devices.
In addition to these quasi-1D excitons, we observe charge-transfer excitons, and their relative energies can be tuned with different screening environments. The easy tunability of the nature of the lowest-energy exciton with external screening is another unique strength of our proposed proximity-induced moir\'e platform. It could be further pushed to explore the adiabatic evolution of optically active, Wannier excitons into potentially long-lived, dark charge-transfer excitons as quantum memories. Overall, our large computed exciton splitting without an external magnetic field provides an accessible two-level system for quantum applications.

\begin{acknowledgement}
We acknowledge helpful discussions with Andrew J. Mannix, Tony F. Heinz, and Allan H. MacDonald.
This work was supported by the U.S. Department of Energy (DOE), Office of Science, Basic Energy Sciences (BES), under award DE-SC0021984.
This work uses codes developed and supported by the Center for Computational Study of Excited State Phenomena in Energy Materials (C2SEPEM), funded by the DOE Office of Science under Contract No. DE-AC02-05CH11231 at the Lawrence Berkeley National Laboratory, as part of the BES Computational Materials Science (CMS) Program.
Support for the computational model for the substrate screening was provided by the National Science Foundation (NSF) through grant no. DMR-2103721.
This research used computational resources at the Oak Ridge Leadership Computing Facility, a DOE Office of Science User Facility supported under Contract No. DE-AC05-00OR22725 for the excited-state calculations, and the Texas Advanced Computing Center at The University of Texas at Austin, funded by the National Science Foundation award 1818253, through allocation DMR21077, for mean-field calculations.
\end{acknowledgement}

\bibliography{ref25.bib}

\providecommand{\latin}[1]{#1}
\makeatletter
\providecommand{\doi}
  {\begingroup\let\do\@makeother\dospecials
  \catcode`\{=1 \catcode`\}=2 \doi@aux}
\providecommand{\doi@aux}[1]{\endgroup\texttt{#1}}
\makeatother
\providecommand*\mcitethebibliography{\thebibliography}
\csname @ifundefined\endcsname{endmcitethebibliography}  {\let\endmcitethebibliography\endthebibliography}{}
\begin{mcitethebibliography}{12}
\providecommand*\natexlab[1]{#1}
\providecommand*\mciteSetBstSublistMode[1]{}
\providecommand*\mciteSetBstMaxWidthForm[2]{}
\providecommand*\mciteBstWouldAddEndPuncttrue
  {\def\EndOfBibitem{\unskip.}}
\providecommand*\mciteBstWouldAddEndPunctfalse
  {\let\EndOfBibitem\relax}
\providecommand*\mciteSetBstMidEndSepPunct[3]{}
\providecommand*\mciteSetBstSublistLabelBeginEnd[3]{}
\providecommand*\EndOfBibitem{}
\mciteSetBstSublistMode{f}
\mciteSetBstMaxWidthForm{subitem}{(\alph{mcitesubitemcount})}
\mciteSetBstSublistLabelBeginEnd
  {\mcitemaxwidthsubitemform\space}
  {\relax}
  {\relax}

\bibitem[Plimpton(1995)]{plimpton1995fast}
Plimpton,~S. Fast parallel algorithms for short-range molecular dynamics. \emph{Journal of computational physics} \textbf{1995}, \emph{117}, 1--19\relax
\mciteBstWouldAddEndPuncttrue
\mciteSetBstMidEndSepPunct{\mcitedefaultmidpunct}
{\mcitedefaultendpunct}{\mcitedefaultseppunct}\relax
\EndOfBibitem
\bibitem[Tersoff(1988)]{tersoff1988new}
Tersoff,~J. New empirical approach for the structure and energy of covalent systems. \emph{Physical review B} \textbf{1988}, \emph{37}, 6991\relax
\mciteBstWouldAddEndPuncttrue
\mciteSetBstMidEndSepPunct{\mcitedefaultmidpunct}
{\mcitedefaultendpunct}{\mcitedefaultseppunct}\relax
\EndOfBibitem
\bibitem[Leven \latin{et~al.}(2014)Leven, Azuri, Kronik, and Hod]{leven2014inter}
Leven,~I.; Azuri,~I.; Kronik,~L.; Hod,~O. Inter-layer potential for hexagonal boron nitride. \emph{The Journal of chemical physics} \textbf{2014}, \emph{140}\relax
\mciteBstWouldAddEndPuncttrue
\mciteSetBstMidEndSepPunct{\mcitedefaultmidpunct}
{\mcitedefaultendpunct}{\mcitedefaultseppunct}\relax
\EndOfBibitem
\bibitem[Maaravi \latin{et~al.}(2017)Maaravi, Leven, Azuri, Kronik, and Hod]{maaravi2017interlayer}
Maaravi,~T.; Leven,~I.; Azuri,~I.; Kronik,~L.; Hod,~O. Interlayer potential for homogeneous graphene and hexagonal boron nitride systems: reparametrization for many-body dispersion effects. \emph{The Journal of Physical Chemistry C} \textbf{2017}, \emph{121}, 22826--22835\relax
\mciteBstWouldAddEndPuncttrue
\mciteSetBstMidEndSepPunct{\mcitedefaultmidpunct}
{\mcitedefaultendpunct}{\mcitedefaultseppunct}\relax
\EndOfBibitem
\bibitem[Giannozzi \latin{et~al.}(2009)Giannozzi, Baroni, Bonini, Calandra, Car, Cavazzoni, Ceresoli, Chiarotti, Cococcioni, Dabo, \latin{et~al.} others]{giannozzi2009quantum}
Giannozzi,~P.; Baroni,~S.; Bonini,~N.; Calandra,~M.; Car,~R.; Cavazzoni,~C.; Ceresoli,~D.; Chiarotti,~G.~L.; Cococcioni,~M.; Dabo,~I.; others QUANTUM ESPRESSO: a modular and open-source software project for quantum simulations of materials. \emph{Journal of physics: Condensed matter} \textbf{2009}, \emph{21}, 395502\relax
\mciteBstWouldAddEndPuncttrue
\mciteSetBstMidEndSepPunct{\mcitedefaultmidpunct}
{\mcitedefaultendpunct}{\mcitedefaultseppunct}\relax
\EndOfBibitem
\bibitem[Perdew \latin{et~al.}(1996)Perdew, Burke, and Ernzerhof]{perdew1996generalized}
Perdew,~J.~P.; Burke,~K.; Ernzerhof,~M. Generalized gradient approximation made simple. \emph{Physical review letters} \textbf{1996}, \emph{77}, 3865\relax
\mciteBstWouldAddEndPuncttrue
\mciteSetBstMidEndSepPunct{\mcitedefaultmidpunct}
{\mcitedefaultendpunct}{\mcitedefaultseppunct}\relax
\EndOfBibitem
\bibitem[Deslippe \latin{et~al.}(2012)Deslippe, Samsonidze, Strubbe, Jain, Cohen, and Louie]{deslippe2012berkeleygw}
Deslippe,~J.; Samsonidze,~G.; Strubbe,~D.~A.; Jain,~M.; Cohen,~M.~L.; Louie,~S.~G. BerkeleyGW: A massively parallel computer package for the calculation of the quasiparticle and optical properties of materials and nanostructures. \emph{Computer Physics Communications} \textbf{2012}, \emph{183}, 1269--1289\relax
\mciteBstWouldAddEndPuncttrue
\mciteSetBstMidEndSepPunct{\mcitedefaultmidpunct}
{\mcitedefaultendpunct}{\mcitedefaultseppunct}\relax
\EndOfBibitem
\bibitem[Altman \latin{et~al.}(2024)Altman, Kundu, and da~Jornada]{altman2024mixed}
Altman,~A.~R.; Kundu,~S.; da~Jornada,~F.~H. Mixed Stochastic-Deterministic Approach for Many-Body Perturbation Theory Calculations. \emph{Physical Review Letters} \textbf{2024}, \emph{132}, 086401\relax
\mciteBstWouldAddEndPuncttrue
\mciteSetBstMidEndSepPunct{\mcitedefaultmidpunct}
{\mcitedefaultendpunct}{\mcitedefaultseppunct}\relax
\EndOfBibitem
\bibitem[Ismail-Beigi(2006)]{ismail2006truncation}
Ismail-Beigi,~S. Truncation of periodic image interactions for confined systems. \emph{Physical Review B—Condensed Matter and Materials Physics} \textbf{2006}, \emph{73}, 233103\relax
\mciteBstWouldAddEndPuncttrue
\mciteSetBstMidEndSepPunct{\mcitedefaultmidpunct}
{\mcitedefaultendpunct}{\mcitedefaultseppunct}\relax
\EndOfBibitem
\bibitem[Keldysh(2024)]{keldysh2024coulomb}
Keldysh,~L. \emph{SELECTED PAPERS OF LEONID V KELDYSH}; World Scientific, 2024; pp 155--158\relax
\mciteBstWouldAddEndPuncttrue
\mciteSetBstMidEndSepPunct{\mcitedefaultmidpunct}
{\mcitedefaultendpunct}{\mcitedefaultseppunct}\relax
\EndOfBibitem
\bibitem[Zhao \latin{et~al.}(2021)Zhao, Xiao, and Yao]{zhao2021universal}
Zhao,~P.; Xiao,~C.; Yao,~W. Universal superlattice potential for 2D materials from twisted interface inside h-BN substrate. \emph{npj 2D Materials and Applications} \textbf{2021}, \emph{5}, 38\relax
\mciteBstWouldAddEndPuncttrue
\mciteSetBstMidEndSepPunct{\mcitedefaultmidpunct}
{\mcitedefaultendpunct}{\mcitedefaultseppunct}\relax
\EndOfBibitem
\end{mcitethebibliography}


\providecommand{\latin}[1]{#1}
\makeatletter
\providecommand{\doi}
  {\begingroup\let\do\@makeother\dospecials
  \catcode`\{=1 \catcode`\}=2 \doi@aux}
\providecommand{\doi@aux}[1]{\endgroup\texttt{#1}}
\makeatother
\providecommand*\mcitethebibliography{\thebibliography}
\csname @ifundefined\endcsname{endmcitethebibliography}  {\let\endmcitethebibliography\endthebibliography}{}
\begin{mcitethebibliography}{50}
\providecommand*\natexlab[1]{#1}
\providecommand*\mciteSetBstSublistMode[1]{}
\providecommand*\mciteSetBstMaxWidthForm[2]{}
\providecommand*\mciteBstWouldAddEndPuncttrue
  {\def\EndOfBibitem{\unskip.}}
\providecommand*\mciteBstWouldAddEndPunctfalse
  {\let\EndOfBibitem\relax}
\providecommand*\mciteSetBstMidEndSepPunct[3]{}
\providecommand*\mciteSetBstSublistLabelBeginEnd[3]{}
\providecommand*\EndOfBibitem{}
\mciteSetBstSublistMode{f}
\mciteSetBstMaxWidthForm{subitem}{(\alph{mcitesubitemcount})}
\mciteSetBstSublistLabelBeginEnd
  {\mcitemaxwidthsubitemform\space}
  {\relax}
  {\relax}

\bibitem[Qiu \latin{et~al.}(2013)Qiu, Da~Jornada, and Louie]{qiu2013optical}
Qiu,~D.~Y.; Da~Jornada,~F.~H.; Louie,~S.~G. Optical spectrum of MoS 2: many-body effects and diversity of exciton states. \emph{Physical review letters} \textbf{2013}, \emph{111}, 216805\relax
\mciteBstWouldAddEndPuncttrue
\mciteSetBstMidEndSepPunct{\mcitedefaultmidpunct}
{\mcitedefaultendpunct}{\mcitedefaultseppunct}\relax
\EndOfBibitem
\bibitem[Chernikov \latin{et~al.}(2014)Chernikov, Berkelbach, Hill, Rigosi, Li, Aslan, Reichman, Hybertsen, and Heinz]{chernikov2014exciton}
Chernikov,~A.; Berkelbach,~T.~C.; Hill,~H.~M.; Rigosi,~A.; Li,~Y.; Aslan,~B.; Reichman,~D.~R.; Hybertsen,~M.~S.; Heinz,~T.~F. Exciton binding energy and nonhydrogenic Rydberg series in monolayer WS 2. \emph{Physical review letters} \textbf{2014}, \emph{113}, 076802\relax
\mciteBstWouldAddEndPuncttrue
\mciteSetBstMidEndSepPunct{\mcitedefaultmidpunct}
{\mcitedefaultendpunct}{\mcitedefaultseppunct}\relax
\EndOfBibitem
\bibitem[Wang \latin{et~al.}(2018)Wang, Chernikov, Glazov, Heinz, Marie, Amand, and Urbaszek]{wang2018colloquium}
Wang,~G.; Chernikov,~A.; Glazov,~M.~M.; Heinz,~T.~F.; Marie,~X.; Amand,~T.; Urbaszek,~B. Colloquium: Excitons in atomically thin transition metal dichalcogenides. \emph{Reviews of Modern Physics} \textbf{2018}, \emph{90}, 021001\relax
\mciteBstWouldAddEndPuncttrue
\mciteSetBstMidEndSepPunct{\mcitedefaultmidpunct}
{\mcitedefaultendpunct}{\mcitedefaultseppunct}\relax
\EndOfBibitem
\bibitem[Yu and Wu(2014)Yu, and Wu]{yu2014valley}
Yu,~T.; Wu,~M. Valley depolarization due to intervalley and intravalley electron-hole exchange interactions in monolayer MoS 2. \emph{Physical Review B} \textbf{2014}, \emph{89}, 205303\relax
\mciteBstWouldAddEndPuncttrue
\mciteSetBstMidEndSepPunct{\mcitedefaultmidpunct}
{\mcitedefaultendpunct}{\mcitedefaultseppunct}\relax
\EndOfBibitem
\bibitem[Glazov \latin{et~al.}(2015)Glazov, Ivchenko, Wang, Amand, Marie, Urbaszek, and Liu]{glazov2015spin}
Glazov,~M.~M.; Ivchenko,~E.~L.; Wang,~G.; Amand,~T.; Marie,~X.; Urbaszek,~B.; Liu,~B. Spin and valley dynamics of excitons in transition metal dichalcogenide monolayers. \emph{physica status solidi (b)} \textbf{2015}, \emph{252}, 2349--2362\relax
\mciteBstWouldAddEndPuncttrue
\mciteSetBstMidEndSepPunct{\mcitedefaultmidpunct}
{\mcitedefaultendpunct}{\mcitedefaultseppunct}\relax
\EndOfBibitem
\bibitem[Kioseoglou \latin{et~al.}(2016)Kioseoglou, Hanbicki, Currie, Friedman, and Jonker]{kioseoglou2016optical}
Kioseoglou,~G.; Hanbicki,~A.~T.; Currie,~M.; Friedman,~A.~L.; Jonker,~B.~T. Optical polarization and intervalley scattering in single layers of MoS2 and MoSe2. \emph{Scientific reports} \textbf{2016}, \emph{6}, 25041\relax
\mciteBstWouldAddEndPuncttrue
\mciteSetBstMidEndSepPunct{\mcitedefaultmidpunct}
{\mcitedefaultendpunct}{\mcitedefaultseppunct}\relax
\EndOfBibitem
\bibitem[Lin \latin{et~al.}(2022)Lin, Liu, Wang, Xu, Chen, Duan, and Monserrat]{lin2022phonon}
Lin,~Z.; Liu,~Y.; Wang,~Z.; Xu,~S.; Chen,~S.; Duan,~W.; Monserrat,~B. Phonon-limited valley polarization in transition-metal dichalcogenides. \emph{Physical review letters} \textbf{2022}, \emph{129}, 027401\relax
\mciteBstWouldAddEndPuncttrue
\mciteSetBstMidEndSepPunct{\mcitedefaultmidpunct}
{\mcitedefaultendpunct}{\mcitedefaultseppunct}\relax
\EndOfBibitem
\bibitem[Jiang \latin{et~al.}(2021)Jiang, Zheng, Lan, Saidi, Ren, and Zhao]{jiang2021real}
Jiang,~X.; Zheng,~Q.; Lan,~Z.; Saidi,~W.~A.; Ren,~X.; Zhao,~J. Real-time GW-BSE investigations on spin-valley exciton dynamics in monolayer transition metal dichalcogenide. \emph{Science Advances} \textbf{2021}, \emph{7}, eabf3759\relax
\mciteBstWouldAddEndPuncttrue
\mciteSetBstMidEndSepPunct{\mcitedefaultmidpunct}
{\mcitedefaultendpunct}{\mcitedefaultseppunct}\relax
\EndOfBibitem
\bibitem[Wang \latin{et~al.}(2024)Wang, Kim, Dong, Shinokita, Watanabe, Taniguchi, and Matsuda]{wang2024quantum}
Wang,~H.; Kim,~H.; Dong,~D.; Shinokita,~K.; Watanabe,~K.; Taniguchi,~T.; Matsuda,~K. Quantum coherence and interference of a single moir{\'e} exciton in nano-fabricated twisted monolayer semiconductor heterobilayers. \emph{Nature Communications} \textbf{2024}, \emph{15}, 4905\relax
\mciteBstWouldAddEndPuncttrue
\mciteSetBstMidEndSepPunct{\mcitedefaultmidpunct}
{\mcitedefaultendpunct}{\mcitedefaultseppunct}\relax
\EndOfBibitem
\bibitem[Durmu{\c{s}} \latin{et~al.}(2023)Durmu{\c{s}}, Demiralay, Khan, Atalay, and Sarpkaya]{durmucs2023prolonged}
Durmu{\c{s}},~M.~A.; Demiralay,~K.; Khan,~M.~M.; Atalay,~{\c{S}}.~E.; Sarpkaya,~I. Prolonged dephasing time of ensemble of moir{\'e}-trapped interlayer excitons in WSe2-MoSe2 heterobilayers. \emph{npj 2D Materials and Applications} \textbf{2023}, \emph{7}, 65\relax
\mciteBstWouldAddEndPuncttrue
\mciteSetBstMidEndSepPunct{\mcitedefaultmidpunct}
{\mcitedefaultendpunct}{\mcitedefaultseppunct}\relax
\EndOfBibitem
\bibitem[Tran \latin{et~al.}(2019)Tran, Moody, Wu, Lu, Choi, Kim, Rai, Sanchez, Quan, Singh, \latin{et~al.} others]{tran2019evidence}
Tran,~K.; Moody,~G.; Wu,~F.; Lu,~X.; Choi,~J.; Kim,~K.; Rai,~A.; Sanchez,~D.~A.; Quan,~J.; Singh,~A.; others Evidence for moir{\'e} excitons in van der Waals heterostructures. \emph{Nature} \textbf{2019}, \emph{567}, 71--75\relax
\mciteBstWouldAddEndPuncttrue
\mciteSetBstMidEndSepPunct{\mcitedefaultmidpunct}
{\mcitedefaultendpunct}{\mcitedefaultseppunct}\relax
\EndOfBibitem
\bibitem[Seyler \latin{et~al.}(2019)Seyler, Rivera, Yu, Wilson, Ray, Mandrus, Yan, Yao, and Xu]{seyler2019signatures}
Seyler,~K.~L.; Rivera,~P.; Yu,~H.; Wilson,~N.~P.; Ray,~E.~L.; Mandrus,~D.~G.; Yan,~J.; Yao,~W.; Xu,~X. Signatures of moir{\'e}-trapped valley excitons in MoSe2/WSe2 heterobilayers. \emph{Nature} \textbf{2019}, \emph{567}, 66--70\relax
\mciteBstWouldAddEndPuncttrue
\mciteSetBstMidEndSepPunct{\mcitedefaultmidpunct}
{\mcitedefaultendpunct}{\mcitedefaultseppunct}\relax
\EndOfBibitem
\bibitem[Kundu \latin{et~al.}(2023)Kundu, Amit, Krishnamurthy, Jain, and Refaely-Abramson]{kundu2023exciton}
Kundu,~S.; Amit,~T.; Krishnamurthy,~H.; Jain,~M.; Refaely-Abramson,~S. Exciton fine structure in twisted transition metal dichalcogenide heterostructures. \emph{npj Computational Materials} \textbf{2023}, \emph{9}, 186\relax
\mciteBstWouldAddEndPuncttrue
\mciteSetBstMidEndSepPunct{\mcitedefaultmidpunct}
{\mcitedefaultendpunct}{\mcitedefaultseppunct}\relax
\EndOfBibitem
\bibitem[Mak and Shan(2022)Mak, and Shan]{mak2022semiconductor}
Mak,~K.~F.; Shan,~J. Semiconductor moir{\'e} materials. \emph{Nature Nanotechnology} \textbf{2022}, \emph{17}, 686--695\relax
\mciteBstWouldAddEndPuncttrue
\mciteSetBstMidEndSepPunct{\mcitedefaultmidpunct}
{\mcitedefaultendpunct}{\mcitedefaultseppunct}\relax
\EndOfBibitem
\bibitem[Huang \latin{et~al.}(2022)Huang, Choi, Shih, and Li]{huang2022excitons}
Huang,~D.; Choi,~J.; Shih,~C.-K.; Li,~X. Excitons in semiconductor moir{\'e} superlattices. \emph{Nature nanotechnology} \textbf{2022}, \emph{17}, 227--238\relax
\mciteBstWouldAddEndPuncttrue
\mciteSetBstMidEndSepPunct{\mcitedefaultmidpunct}
{\mcitedefaultendpunct}{\mcitedefaultseppunct}\relax
\EndOfBibitem
\bibitem[Wilson \latin{et~al.}(2021)Wilson, Yao, Shan, and Xu]{wilson2021excitons}
Wilson,~N.~P.; Yao,~W.; Shan,~J.; Xu,~X. Excitons and emergent quantum phenomena in stacked 2D semiconductors. \emph{Nature} \textbf{2021}, \emph{599}, 383--392\relax
\mciteBstWouldAddEndPuncttrue
\mciteSetBstMidEndSepPunct{\mcitedefaultmidpunct}
{\mcitedefaultendpunct}{\mcitedefaultseppunct}\relax
\EndOfBibitem
\bibitem[Thureja \latin{et~al.}(2022)Thureja, Imamoglu, Smole{\'n}ski, Amelio, Popert, Chervy, Lu, Liu, Barmak, Watanabe, \latin{et~al.} others]{thureja2022electrically}
Thureja,~D.; Imamoglu,~A.; Smole{\'n}ski,~T.; Amelio,~I.; Popert,~A.; Chervy,~T.; Lu,~X.; Liu,~S.; Barmak,~K.; Watanabe,~K.; others Electrically tunable quantum confinement of neutral excitons. \emph{Nature} \textbf{2022}, \emph{606}, 298--304\relax
\mciteBstWouldAddEndPuncttrue
\mciteSetBstMidEndSepPunct{\mcitedefaultmidpunct}
{\mcitedefaultendpunct}{\mcitedefaultseppunct}\relax
\EndOfBibitem
\bibitem[Heithoff \latin{et~al.}(2023)Heithoff, Moreno, Torre, Feuer, Purser, Andolina, Calajo, Watanabe, Taniguchi, Kara, \latin{et~al.} others]{heithoff2023valley}
Heithoff,~M.; Moreno,~{\'A}.; Torre,~I.; Feuer,~M.~S.; Purser,~C.~M.; Andolina,~G.~M.; Calajo,~G.; Watanabe,~K.; Taniguchi,~T.; Kara,~D.; others Valley-hybridized gate-tunable 1D exciton confinement in MoSe2. \emph{arXiv preprint arXiv:2311.05299} \textbf{2023}, \relax
\mciteBstWouldAddEndPunctfalse
\mciteSetBstMidEndSepPunct{\mcitedefaultmidpunct}
{}{\mcitedefaultseppunct}\relax
\EndOfBibitem
\bibitem[Hu \latin{et~al.}(2024)Hu, Lorchat, Chen, Watanabe, Taniguchi, Heinz, Murthy, and Chervy]{hu2024quantum}
Hu,~J.; Lorchat,~E.; Chen,~X.; Watanabe,~K.; Taniguchi,~T.; Heinz,~T.~F.; Murthy,~P.~A.; Chervy,~T. Quantum control of exciton wave functions in 2D semiconductors. \emph{Science Advances} \textbf{2024}, \emph{10}, eadk6369\relax
\mciteBstWouldAddEndPuncttrue
\mciteSetBstMidEndSepPunct{\mcitedefaultmidpunct}
{\mcitedefaultendpunct}{\mcitedefaultseppunct}\relax
\EndOfBibitem
\bibitem[Wang \latin{et~al.}(2021)Wang, Maisch, Tang, Zhao, Yang, Joos, Portalupi, Michler, and Smet]{wang2021highly}
Wang,~Q.; Maisch,~J.; Tang,~F.; Zhao,~D.; Yang,~S.; Joos,~R.; Portalupi,~S.~L.; Michler,~P.; Smet,~J.~H. Highly polarized single photons from strain-induced quasi-1D localized excitons in WSe2. \emph{Nano letters} \textbf{2021}, \emph{21}, 7175--7182\relax
\mciteBstWouldAddEndPuncttrue
\mciteSetBstMidEndSepPunct{\mcitedefaultmidpunct}
{\mcitedefaultendpunct}{\mcitedefaultseppunct}\relax
\EndOfBibitem
\bibitem[He \latin{et~al.}(2015)He, Clark, Schaibley, He, Chen, Wei, Ding, Zhang, Yao, Xu, \latin{et~al.} others]{he2015single}
He,~Y.-M.; Clark,~G.; Schaibley,~J.~R.; He,~Y.; Chen,~M.-C.; Wei,~Y.-J.; Ding,~X.; Zhang,~Q.; Yao,~W.; Xu,~X.; others Single quantum emitters in monolayer semiconductors. \emph{Nature nanotechnology} \textbf{2015}, \emph{10}, 497--502\relax
\mciteBstWouldAddEndPuncttrue
\mciteSetBstMidEndSepPunct{\mcitedefaultmidpunct}
{\mcitedefaultendpunct}{\mcitedefaultseppunct}\relax
\EndOfBibitem
\bibitem[Chakraborty \latin{et~al.}(2015)Chakraborty, Kinnischtzke, Goodfellow, Beams, and Vamivakas]{chakraborty2015voltage}
Chakraborty,~C.; Kinnischtzke,~L.; Goodfellow,~K.~M.; Beams,~R.; Vamivakas,~A.~N. Voltage-controlled quantum light from an atomically thin semiconductor. \emph{Nature nanotechnology} \textbf{2015}, \emph{10}, 507--511\relax
\mciteBstWouldAddEndPuncttrue
\mciteSetBstMidEndSepPunct{\mcitedefaultmidpunct}
{\mcitedefaultendpunct}{\mcitedefaultseppunct}\relax
\EndOfBibitem
\bibitem[Srivastava \latin{et~al.}(2015)Srivastava, Sidler, Allain, Lembke, Kis, and Imamo{\u{g}}lu]{srivastava2015optically}
Srivastava,~A.; Sidler,~M.; Allain,~A.~V.; Lembke,~D.~S.; Kis,~A.; Imamo{\u{g}}lu,~A. Optically active quantum dots in monolayer WSe2. \emph{Nature nanotechnology} \textbf{2015}, \emph{10}, 491--496\relax
\mciteBstWouldAddEndPuncttrue
\mciteSetBstMidEndSepPunct{\mcitedefaultmidpunct}
{\mcitedefaultendpunct}{\mcitedefaultseppunct}\relax
\EndOfBibitem
\bibitem[Bai \latin{et~al.}(2020)Bai, Zhou, Wang, Wu, McGilly, Halbertal, Lo, Liu, Ardelean, Rivera, \latin{et~al.} others]{bai2020excitons}
Bai,~Y.; Zhou,~L.; Wang,~J.; Wu,~W.; McGilly,~L.~J.; Halbertal,~D.; Lo,~C. F.~B.; Liu,~F.; Ardelean,~J.; Rivera,~P.; others Excitons in strain-induced one-dimensional moir{\'e} potentials at transition metal dichalcogenide heterojunctions. \emph{Nature Materials} \textbf{2020}, \emph{19}, 1068--1073\relax
\mciteBstWouldAddEndPuncttrue
\mciteSetBstMidEndSepPunct{\mcitedefaultmidpunct}
{\mcitedefaultendpunct}{\mcitedefaultseppunct}\relax
\EndOfBibitem
\bibitem[Yasuda \latin{et~al.}(2021)Yasuda, Wang, Watanabe, Taniguchi, and Jarillo-Herrero]{yasuda2021stacking}
Yasuda,~K.; Wang,~X.; Watanabe,~K.; Taniguchi,~T.; Jarillo-Herrero,~P. Stacking-engineered ferroelectricity in bilayer boron nitride. \emph{Science} \textbf{2021}, \emph{372}, 1458--1462\relax
\mciteBstWouldAddEndPuncttrue
\mciteSetBstMidEndSepPunct{\mcitedefaultmidpunct}
{\mcitedefaultendpunct}{\mcitedefaultseppunct}\relax
\EndOfBibitem
\bibitem[Lv \latin{et~al.}(2022)Lv, Sun, Chen, Taniguchi, Watanabe, Wu, Wang, and Xue]{lv2022spatially}
Lv,~M.; Sun,~X.; Chen,~Y.; Taniguchi,~T.; Watanabe,~K.; Wu,~M.; Wang,~J.; Xue,~J. Spatially resolved polarization manipulation of ferroelectricity in twisted hBN. \emph{Advanced Materials} \textbf{2022}, \emph{34}, 2203990\relax
\mciteBstWouldAddEndPuncttrue
\mciteSetBstMidEndSepPunct{\mcitedefaultmidpunct}
{\mcitedefaultendpunct}{\mcitedefaultseppunct}\relax
\EndOfBibitem
\bibitem[Woods \latin{et~al.}(2021)Woods, Ares, Nevison-Andrews, Holwill, Fabregas, Guinea, Geim, Novoselov, Walet, and Fumagalli]{woods2021charge}
Woods,~C.; Ares,~P.; Nevison-Andrews,~H.; Holwill,~M.; Fabregas,~R.; Guinea,~F.; Geim,~A.; Novoselov,~K.; Walet,~N.; Fumagalli,~L. Charge-polarized interfacial superlattices in marginally twisted hexagonal boron nitride. \emph{Nature communications} \textbf{2021}, \emph{12}, 347\relax
\mciteBstWouldAddEndPuncttrue
\mciteSetBstMidEndSepPunct{\mcitedefaultmidpunct}
{\mcitedefaultendpunct}{\mcitedefaultseppunct}\relax
\EndOfBibitem
\bibitem[Zhao \latin{et~al.}(2021)Zhao, Xiao, and Yao]{zhao2021universal}
Zhao,~P.; Xiao,~C.; Yao,~W. Universal superlattice potential for 2D materials from twisted interface inside h-BN substrate. \emph{npj 2D Materials and Applications} \textbf{2021}, \emph{5}, 38\relax
\mciteBstWouldAddEndPuncttrue
\mciteSetBstMidEndSepPunct{\mcitedefaultmidpunct}
{\mcitedefaultendpunct}{\mcitedefaultseppunct}\relax
\EndOfBibitem
\bibitem[Zhou \latin{et~al.}(2022)Zhou, Kotiuga, and Darancet]{zhou2022analyticaltheorynearfieldelectrostatic}
Zhou,~Q.; Kotiuga,~M.; Darancet,~P. Analytical Theory of Near-Field Electrostatic Effects in Two-Dimensional Materials and van der Waals Heterojunctions. 2022; \url{https://arxiv.org/abs/2205.04606}\relax
\mciteBstWouldAddEndPuncttrue
\mciteSetBstMidEndSepPunct{\mcitedefaultmidpunct}
{\mcitedefaultendpunct}{\mcitedefaultseppunct}\relax
\EndOfBibitem
\bibitem[Kim \latin{et~al.}(2024)Kim, Dominguez, Mayorga-Luna, Ye, Embley, Tan, Ni, Liu, Ford, Gao, \latin{et~al.} others]{kim2024electrostatic}
Kim,~D.~S.; Dominguez,~R.~C.; Mayorga-Luna,~R.; Ye,~D.; Embley,~J.; Tan,~T.; Ni,~Y.; Liu,~Z.; Ford,~M.; Gao,~F.~Y.; others Electrostatic moir{\'e} potential from twisted hexagonal boron nitride layers. \emph{Nature materials} \textbf{2024}, \emph{23}, 65--70\relax
\mciteBstWouldAddEndPuncttrue
\mciteSetBstMidEndSepPunct{\mcitedefaultmidpunct}
{\mcitedefaultendpunct}{\mcitedefaultseppunct}\relax
\EndOfBibitem
\bibitem[Kim \latin{et~al.}(2024)Kim, Xiao, Dominguez, Liu, Abudayyeh, Lee, Mayorga-Luna, Kim, Watanabe, Taniguchi, \latin{et~al.} others]{kim2024harnessing}
Kim,~D.~S.; Xiao,~C.; Dominguez,~R.~C.; Liu,~Z.; Abudayyeh,~H.; Lee,~K.; Mayorga-Luna,~R.; Kim,~H.; Watanabe,~K.; Taniguchi,~T.; others Harnessing moir$\backslash$'e ferroelectricity to modulate semiconductor monolayer light emission. \emph{arXiv preprint arXiv:2405.11159} \textbf{2024}, \relax
\mciteBstWouldAddEndPunctfalse
\mciteSetBstMidEndSepPunct{\mcitedefaultmidpunct}
{}{\mcitedefaultseppunct}\relax
\EndOfBibitem
\bibitem[Gu \latin{et~al.}(2025)Gu, Zhang, Felsenfeld, Gao, Ma, Park, Jang, Taniguchi, Watanabe, and Zhou]{gu2025quantum}
Gu,~L.; Zhang,~L.; Felsenfeld,~S.; Gao,~B.; Ma,~R.; Park,~S.; Jang,~H.; Taniguchi,~T.; Watanabe,~K.; Zhou,~Y. Quantum confining excitons with an electrostatic moir{\'e} superlattice. \emph{Physical Review Letters} \textbf{2025}, \emph{135}, 026901\relax
\mciteBstWouldAddEndPuncttrue
\mciteSetBstMidEndSepPunct{\mcitedefaultmidpunct}
{\mcitedefaultendpunct}{\mcitedefaultseppunct}\relax
\EndOfBibitem
\bibitem[Li and Wu(2017)Li, and Wu]{li2017binary}
Li,~L.; Wu,~M. Binary compound bilayer and multilayer with vertical polarizations: two-dimensional ferroelectrics, multiferroics, and nanogenerators. \emph{ACS nano} \textbf{2017}, \emph{11}, 6382--6388\relax
\mciteBstWouldAddEndPuncttrue
\mciteSetBstMidEndSepPunct{\mcitedefaultmidpunct}
{\mcitedefaultendpunct}{\mcitedefaultseppunct}\relax
\EndOfBibitem
\bibitem[Cho \latin{et~al.}(2024)Cho, Datta, Han, Chand, Adak, Yu, Li, Watanabe, Taniguchi, Hone, \latin{et~al.} others]{cho2024moire}
Cho,~M.; Datta,~B.; Han,~K.; Chand,~S.~B.; Adak,~P.~C.; Yu,~S.; Li,~F.; Watanabe,~K.; Taniguchi,~T.; Hone,~J.; others Moir{\'e} exciton polaron engineering via twisted hBN. \emph{Nano Letters} \textbf{2024}, \emph{25}, 1381--1388\relax
\mciteBstWouldAddEndPuncttrue
\mciteSetBstMidEndSepPunct{\mcitedefaultmidpunct}
{\mcitedefaultendpunct}{\mcitedefaultseppunct}\relax
\EndOfBibitem
\bibitem[Rohlfing and Louie(2000)Rohlfing, and Louie]{rohlfing2000electron}
Rohlfing,~M.; Louie,~S.~G. Electron-hole excitations and optical spectra from first principles. \emph{Physical Review B} \textbf{2000}, \emph{62}, 4927\relax
\mciteBstWouldAddEndPuncttrue
\mciteSetBstMidEndSepPunct{\mcitedefaultmidpunct}
{\mcitedefaultendpunct}{\mcitedefaultseppunct}\relax
\EndOfBibitem
\bibitem[Deslippe \latin{et~al.}(2012)Deslippe, Samsonidze, Strubbe, Jain, Cohen, and Louie]{deslippe2012berkeleygw}
Deslippe,~J.; Samsonidze,~G.; Strubbe,~D.~A.; Jain,~M.; Cohen,~M.~L.; Louie,~S.~G. BerkeleyGW: A massively parallel computer package for the calculation of the quasiparticle and optical properties of materials and nanostructures. \emph{Computer Physics Communications} \textbf{2012}, \emph{183}, 1269--1289\relax
\mciteBstWouldAddEndPuncttrue
\mciteSetBstMidEndSepPunct{\mcitedefaultmidpunct}
{\mcitedefaultendpunct}{\mcitedefaultseppunct}\relax
\EndOfBibitem
\bibitem[Qiu \latin{et~al.}(2015)Qiu, Cao, and Louie]{qiu2015nonanalyticity}
Qiu,~D.~Y.; Cao,~T.; Louie,~S.~G. Nonanalyticity, valley quantum phases, and lightlike exciton dispersion in monolayer transition metal dichalcogenides: Theory and first-principles calculations. \emph{Physical review letters} \textbf{2015}, \emph{115}, 176801\relax
\mciteBstWouldAddEndPuncttrue
\mciteSetBstMidEndSepPunct{\mcitedefaultmidpunct}
{\mcitedefaultendpunct}{\mcitedefaultseppunct}\relax
\EndOfBibitem
\bibitem[Yu \latin{et~al.}(2014)Yu, Liu, Gong, Xu, and Yao]{yu2014dirac}
Yu,~H.; Liu,~G.-B.; Gong,~P.; Xu,~X.; Yao,~W. Dirac cones and Dirac saddle points of bright excitons in monolayer transition metal dichalcogenides. \emph{Nature communications} \textbf{2014}, \emph{5}, 3876\relax
\mciteBstWouldAddEndPuncttrue
\mciteSetBstMidEndSepPunct{\mcitedefaultmidpunct}
{\mcitedefaultendpunct}{\mcitedefaultseppunct}\relax
\EndOfBibitem
\bibitem[Frauni{\'e} \latin{et~al.}(2023)Frauni{\'e}, Jamil, Kantelberg, Roux, Petit, Lepleux, Pacheco, Watanabe, Taniguchi, Jacques, \latin{et~al.} others]{fraunie2023electron}
Frauni{\'e},~J.; Jamil,~R.; Kantelberg,~R.; Roux,~S.; Petit,~L.; Lepleux,~E.; Pacheco,~L.; Watanabe,~K.; Taniguchi,~T.; Jacques,~V.; others Electron and hole doping of monolayer WSe2 induced by twisted ferroelectric hexagonal boron nitride. \emph{Physical Review Materials} \textbf{2023}, \emph{7}, L121002\relax
\mciteBstWouldAddEndPuncttrue
\mciteSetBstMidEndSepPunct{\mcitedefaultmidpunct}
{\mcitedefaultendpunct}{\mcitedefaultseppunct}\relax
\EndOfBibitem
\bibitem[Bennett \latin{et~al.}(2023)Bennett, Chaudhary, Slager, Bousquet, and Ghosez]{bennett2023polar}
Bennett,~D.; Chaudhary,~G.; Slager,~R.-J.; Bousquet,~E.; Ghosez,~P. Polar meron-antimeron networks in strained and twisted bilayers. \emph{Nature Communications} \textbf{2023}, \emph{14}, 1629\relax
\mciteBstWouldAddEndPuncttrue
\mciteSetBstMidEndSepPunct{\mcitedefaultmidpunct}
{\mcitedefaultendpunct}{\mcitedefaultseppunct}\relax
\EndOfBibitem
\bibitem[Plimpton(1995)]{plimpton1995fast}
Plimpton,~S. Fast parallel algorithms for short-range molecular dynamics. \emph{Journal of computational physics} \textbf{1995}, \emph{117}, 1--19\relax
\mciteBstWouldAddEndPuncttrue
\mciteSetBstMidEndSepPunct{\mcitedefaultmidpunct}
{\mcitedefaultendpunct}{\mcitedefaultseppunct}\relax
\EndOfBibitem
\bibitem[Kohn and Sham(1996)Kohn, and Sham]{kohn1996density}
Kohn,~W.; Sham,~L. Density functional theory. Conference Proceedings-Italian Physical Society. 1996; pp 561--572\relax
\mciteBstWouldAddEndPuncttrue
\mciteSetBstMidEndSepPunct{\mcitedefaultmidpunct}
{\mcitedefaultendpunct}{\mcitedefaultseppunct}\relax
\EndOfBibitem
\bibitem[Giannozzi \latin{et~al.}(2009)Giannozzi, Baroni, Bonini, Calandra, Car, Cavazzoni, Ceresoli, Chiarotti, Cococcioni, Dabo, \latin{et~al.} others]{giannozzi2009quantum}
Giannozzi,~P.; Baroni,~S.; Bonini,~N.; Calandra,~M.; Car,~R.; Cavazzoni,~C.; Ceresoli,~D.; Chiarotti,~G.~L.; Cococcioni,~M.; Dabo,~I.; others QUANTUM ESPRESSO: a modular and open-source software project for quantum simulations of materials. \emph{Journal of physics: Condensed matter} \textbf{2009}, \emph{21}, 395502\relax
\mciteBstWouldAddEndPuncttrue
\mciteSetBstMidEndSepPunct{\mcitedefaultmidpunct}
{\mcitedefaultendpunct}{\mcitedefaultseppunct}\relax
\EndOfBibitem
\bibitem[Altman \latin{et~al.}(2024)Altman, Kundu, and da~Jornada]{altman2024mixed}
Altman,~A.~R.; Kundu,~S.; da~Jornada,~F.~H. Mixed Stochastic-Deterministic Approach for Many-Body Perturbation Theory Calculations. \emph{Physical Review Letters} \textbf{2024}, \emph{132}, 086401\relax
\mciteBstWouldAddEndPuncttrue
\mciteSetBstMidEndSepPunct{\mcitedefaultmidpunct}
{\mcitedefaultendpunct}{\mcitedefaultseppunct}\relax
\EndOfBibitem
\bibitem[Keldysh(2024)]{keldysh2024coulomb}
Keldysh,~L. \emph{SELECTED PAPERS OF LEONID V KELDYSH}; World Scientific, 2024; pp 155--158\relax
\mciteBstWouldAddEndPuncttrue
\mciteSetBstMidEndSepPunct{\mcitedefaultmidpunct}
{\mcitedefaultendpunct}{\mcitedefaultseppunct}\relax
\EndOfBibitem
\bibitem[da~Jornada \latin{et~al.}(2020)da~Jornada, Xian, Rubio, and Louie]{da2020universal}
da~Jornada,~F.~H.; Xian,~L.; Rubio,~A.; Louie,~S.~G. Universal slow plasmons and giant field enhancement in atomically thin quasi-two-dimensional metals. \emph{Nature communications} \textbf{2020}, \emph{11}, 1013\relax
\mciteBstWouldAddEndPuncttrue
\mciteSetBstMidEndSepPunct{\mcitedefaultmidpunct}
{\mcitedefaultendpunct}{\mcitedefaultseppunct}\relax
\EndOfBibitem
\bibitem[Cavalcante \latin{et~al.}(2018)Cavalcante, da~Costa, Farias, Reichman, and Chaves]{cavalcante2018stark}
Cavalcante,~L.; da~Costa,~D.~R.; Farias,~G.; Reichman,~D.; Chaves,~A. Stark shift of excitons and trions in two-dimensional materials. \emph{Physical Review B} \textbf{2018}, \emph{98}, 245309\relax
\mciteBstWouldAddEndPuncttrue
\mciteSetBstMidEndSepPunct{\mcitedefaultmidpunct}
{\mcitedefaultendpunct}{\mcitedefaultseppunct}\relax
\EndOfBibitem
\bibitem[Zhou \latin{et~al.}(2024)Zhou, Gon{\c c}alves, Riminucci, Dhuey, S.~Barnard, Schwartzberg, Garc{\'\i}a~de Abajo, and Weber-Bargioni]{zhou2024probing}
Zhou,~J.; Gon{\c c}alves,~P. A.~D.; Riminucci,~F.; Dhuey,~S.; S.~Barnard,~E.; Schwartzberg,~A.; Garc{\'\i}a~de Abajo,~F.~J.; Weber-Bargioni,~A. Probing plexciton emission from 2D materials on gold nanotrenches. \emph{Nature Communications} \textbf{2024}, \emph{15}, 9583\relax
\mciteBstWouldAddEndPuncttrue
\mciteSetBstMidEndSepPunct{\mcitedefaultmidpunct}
{\mcitedefaultendpunct}{\mcitedefaultseppunct}\relax
\EndOfBibitem
\bibitem[Han \latin{et~al.}(2024)Han, Cho, Kim, Kim, Kim, Park, Yang, Watanabe, Taniguchi, Menon, \latin{et~al.} others]{han2024highly}
Han,~K.; Cho,~M.; Kim,~T.; Kim,~S.~T.; Kim,~S.~H.; Park,~S.~H.; Yang,~S.~M.; Watanabe,~K.; Taniguchi,~T.; Menon,~V.; others Highly tunable moir{\'e} superlattice potentials in twisted hexagonal boron nitrides. \emph{Advanced Science} \textbf{2024}, 2408034\relax
\mciteBstWouldAddEndPuncttrue
\mciteSetBstMidEndSepPunct{\mcitedefaultmidpunct}
{\mcitedefaultendpunct}{\mcitedefaultseppunct}\relax
\EndOfBibitem
\end{mcitethebibliography}
\end{document}


\maketitle

\section{Computational details}

We relax the twisted hBN (t-hBN) using classical molecular simulation as implemented in the LAAMPS package \cite{plimpton1995fast}. The lattice constant used for hBN is 2.5\AA{}. The intralayer potential is represented by Tersoff forcefield \cite{tersoff1988new} and the interlayer interaction is described by Kolmogorov-Crespi potential and Coulomb interaction which takes into account the stacking-dependent charge polarization \cite{leven2014inter,maaravi2017interlayer}. The forces are minimized using the conjugate gradient method till the force on each atom becomes less than $10^{-8}$ eV/\AA{}.
The effective polarization potential map, experienced by the TMD in the t-hBN/TMD heterostructure, is constructed by computing the shift of the band-edge of MoS$_2$ in presence of a bilayer hBN with a particular stacking corresponding to the local stacking in the moir\'e cell. The computation of the effective potential, using multiple TMD-bilayer hBN configurations, is performed using DFT as implemented in Quantum ESPRESSO \cite{giannozzi2009quantum}. Defining the effective polarization in such way automatically captures the screening from the TMD layer.

The electronic structure of the MoS$_2$ supercell, including the effect of the polarization potential arising from the t-hBN is, computed using Quantum Espresso. 
We add the effective external potential $V_p$ to the Kohn-Sham equation (cite) while solving it self-consistently, approximating it to be constant along the out-of-plane $z$ coordinate and only depending on the in-plane coordinates. From linear response theory, we expect MoS$_2$ to partially screen out \emph{changes} in $V_p$, the largest $E_{xy}=\nabla_{xy} V_p$ occurring at the DWs. This leads to an induced in-plane dipole on monolayer MoS$_2$ at regions on top of the DWs. In a supercell arrangement, these dipoles lead to a surface charge density perpendicular to the DW, and hence a finite discontinuity in the electric potential, spuriously decreasing the bandwidth of the effective potential $V_p$. We stress that this effect is spurious from running our calculation in supercell geometry with a finite thickness $L_z$ along the out-of-plane direction. We circumvent this spurious potential induced along the interface by rescaling the external potential by a factor of 7, such that the shifts of the valence band maximum and the conduction band minimum at the center of the domains match those observed when we considered our previous calculations on monolayer MoS$_2$ supported on AB- and BA-stacked bilayer hBN. The scaling factor is consistent with the in-plane dielectric screening of the supercell with $L_z=15$ \AA{}, since only about 40\% of the supercell volume is occupied by the material compared to bulk TMDs. We therefore expect the effective in-plane dielectric screening to be reduced to roughly 40\% of the bulk value.
As a result of the external potential, the local bands of MoS$_2$ at the edge of the supercell, corresponding to the BA stacked hBN, shift to lower energies. The local bands at the center of the supercell shift to higher energies, as expected for a MoS$_2$ stacked on an AB bilayer hBN. In the intermediate regions of the supercell, the band edges continuously bend following the external potential profile. 
As an effect of the scaled external potential, the band gap is reduced by 148 meV.

The lattice constant of MoS$_2$ unit cell is 3.15 \AA{}. A vacuum of 12\AA{} is employed in the out-of-plane direction. We use norm-conserving pseudopotential, and the exchange-correlation functional is approximated by generalized gradient approximation \cite{perdew1996generalized}. The wave functions are expanded on a plane-wave basis with energy up to 40 Ry. We use a $1\times48\times1$ k-grid sampling of the supercell Brillouin zone. Spin-orbit interaction is included. 

We study the excitons in the MoS$_2$ supercell, including the external potential using the BerkeleyGW package \cite{deslippe2012berkeleygw}. We solve the Bethe-Salpeter equation and include 40 valence bands and 56 conduction bands of the supercell spanning 500 meV energy below the valence band maximum and conduction band minimum, respectively. The electron-hole interaction kernels are computed and the excitons are solved on $1\times48\times1$ k-grid.The absorption spectra are plotted with a Gaussian broadening of 10 meV. We employ a stochastic compression scheme for the bands away from the band edges to accelerate the computation of the dielectric matrix of the MoS$_2$ supercell \cite{altman2024mixed}. A truncation scheme is employed to truncate the Coulomb interaction along the out-of-plane direction \cite{ismail2006truncation}. The effect of the external dielectric environment is included by exploiting the additive property of the polarizability matrix ($\chi$) and employing a two-dimensional screening model.

\section{External dielectric environment}

Considering the fact that the bands of MoS$_2$ near the band edges do not hybridize with those of the substrate dielectric, the irreducible polarizability matrix of the total system ($\chi^0_{tot}$) can be written as the addition of the polarizability matrix of individual components: $\chi^0_{tot}= \chi^0_M+\chi^0_{S}$, where $\chi^0_M$ and $\chi^0_S$ represent the irreducible polarizability of MoS$_2$ and the external substrate, such as thick hBN, respectively. 
Correspondingly, the dielectric matrix of the full structure can be written as:
\begin{align}
    \epsilon_{tot} &= 1 - v\chi^0_{tot}\\ \nonumber
    &= 1 - v(\chi^0_M+\chi^0_{S})\\ \nonumber
    &= (1-v\chi^0_{S})- v\chi^0_{M}
\end{align}
Denoting $1-v\chi^0_{S}$ as $\epsilon_S$ (dielectric of the substrate), we can write
\begin{align}
    \epsilon_{tot} &= \epsilon_S - v\chi^0_{M}\\
     \epsilon_{tot}^{-1} &= (\epsilon_S - v\chi^0_{M})^{-1} \\ \nonumber
     &= (1-\epsilon_S^{-1} v\chi^0_{M})^{-1}\epsilon_S^{-1}\\ \nonumber
     &=(1-W_{S}\chi^0_{M})^{-1}\epsilon_S^{-1}\\ \nonumber
     &= \tilde{\epsilon}^{-1}\epsilon_S^{-1}
\end{align}

The polarizability and the dielectric matrix of MoS$_2$ ($\chi^0_{M}$) is calculated using BerkeleyGW package while the dielectric function of the substrate is approximated with a momentum-dependent Keldysh model \cite{keldysh2024coulomb}. To avoid double counting the screening of the MoS$_2$ layer, we set the material screening to be 1 in computing the effective dielectric due to the substrate, as illustrated below:

\begin{align}
    \epsilon_S(q) =\frac{1}{2} sech(\frac{qd}{2}+\eta_1)sech(\frac{qd}{2}+\eta_2)sinh(qd+\eta_1+\eta_2)
    \label{eqn:keldysh1}
\end{align}
where $\eta_i=\frac{1}{2}log{\frac{\epsilon_i+1}{\epsilon_i-1}}+\frac{i\pi}{2}=\tilde{\eta}_i+\frac{i\pi}{2}$ with $\epsilon_i$ being the dielectric constant of the substrate, $d$ is the effective thickness of the material and $q$ is the magnitude of the in-plane momentum. Simplifying the expression and denoting $x=\frac{(q+G_{xy})d}{2}$ where $G_{xy}$ is the in-plane component of the reciprocal lattice vectors and $t=\textrm{tanh}(x)$, Eqn. \ref{eqn:keldysh1} becomes,
\begin{align}
    \epsilon_S(q) =\frac{1}{2}\left[ \frac{\epsilon_1+t}{1+\epsilon_1 t}+ \frac{\epsilon_2+t}{1+\epsilon_2 t}\right]
\end{align}
We use $\epsilon_1$=4.5, $\epsilon_2=1$ for hBN substrate, and $\epsilon_2=\epsilon_1$=4.5 for hBN encapsulation environment.  

\section{Ab initio exciton wavefunctions}

\begin{figure}[h]
    \includegraphics[scale=0.4]{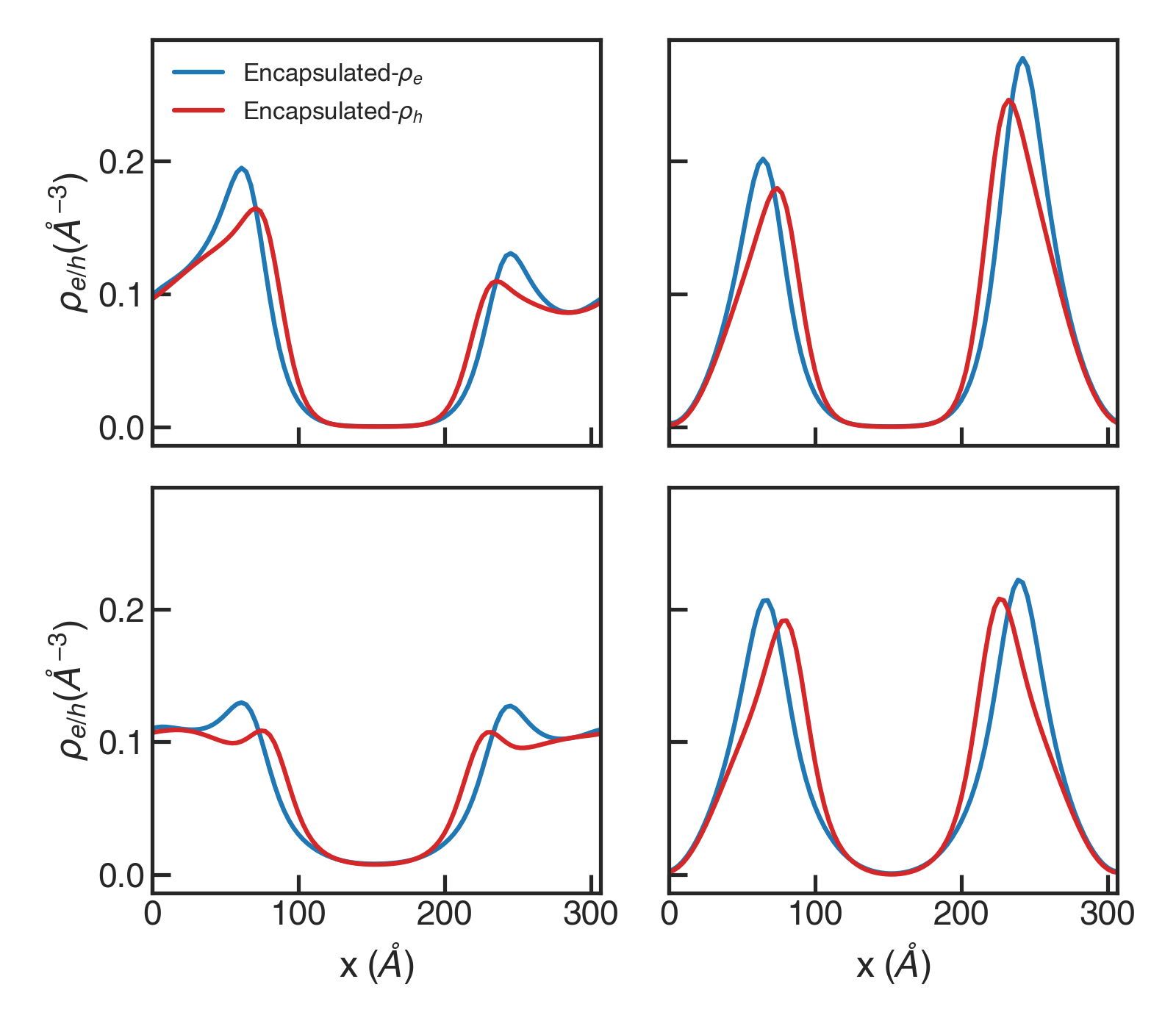}
    \caption{(a)-(b) The electron and hole density distributions while integrating over the other particle are shown in blue and red respectively, for the first two transversely polarized excitons: $X^W_{T0}$ and $X^W_{T1}$  (c)-(d): Same plots for the first two excitons, which couple with the longitudinal component of the photon electric field ($X^W_{L0}$ and $X^W_{L1}$).}
    \label{fig:si_fig1}
\end{figure}

We show the electron (hole) density distribution while integrating over the hole (electron): $\rho_e$ ($\rho_h$) for the first two bright excitons showing polarization along the transverse direction ($X^W_T$) and the longitudinal direction ($X^W_L$) in Fig. \ref{fig:si_fig1}. We observe that $X^W_{T0}$ and $X^W_{T1}$ localize at both the left and right domain wall (DW) regions. The two states are energetically separated by 1.2 meV when the heterostructure is encapsulated by hBN. 
The excitons can be viewed as two excitons localized at the two DW's forming a bonding and anti-bonding combination of them ($X^W_{T0}$ and $X^W_{T1}$). However, we notice that the $X^W_{T0}$ shows finite density at the edge of the supercell but not at the center of it. This behavior is attributed to the difference in effective masses of the electron and hole constituting $X^W_{T0}$. On the other hand, $X^W_{T1}$ is perfectly localized at the DW regions, signifying the anti-bonding nature.
We average over the two states in Fig. 4(b) of the manuscript. 
At higher energies, the kinetic energies of the excitons become larger and hence the longitudianlly polarized excitons ($X^W_{L}$) are more delocalized.

\section{Distance dependence of electric field}

\begin{figure}[h]
    \centering
    \includegraphics[scale=0.5]{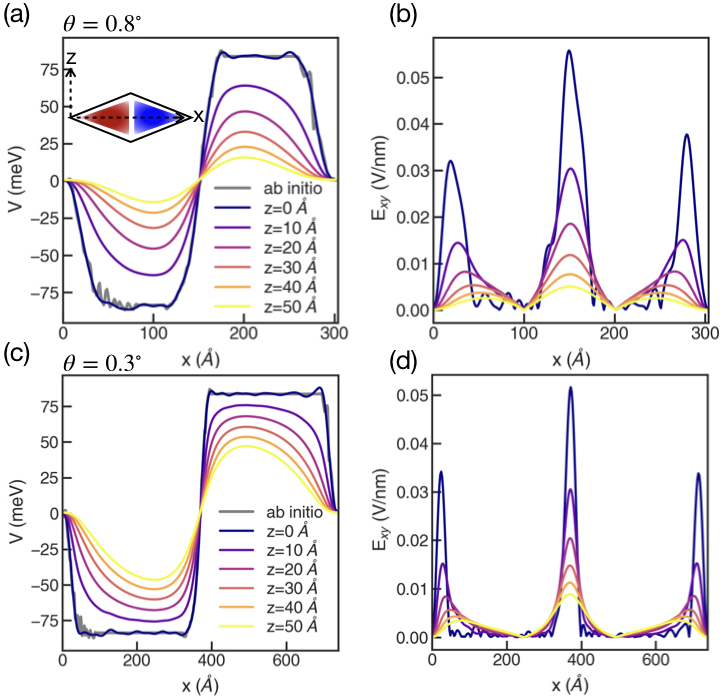}
    \caption{(a): The polarization potential due to a 0.82\si{\degree} t-hBN at different distances along the out-of-plane direction ($z$) from the twisted interface. The $ab initio$ potential ($V_p$) is shown in grey color. The inset depicts the axis orientations. (b): The out-of-plane component of the resulting electric field for different $z$ values for $\theta=0.82\si{\degree}$. (c)-(d): Same as (a) and (b) for $\theta=0.33\si{\degree}$.}
    \label{fig:si_fig3}
\end{figure}

The in-plane polarization potential at the twisted interface of t-hBN ($V_P(\textbf{R},z=0)\equiv V_P(\textbf{R})$) can be expanded in terms of the moir\'e reciprocal lattice vectors (\textbf{G}): $V_P(\textbf{R})=\sum_{\textbf{G}}V_{\textbf{G}}e^{i\textbf{G}.\textbf{R}}$, where $R$ is the in-plane position in the moir\'e supercell. We found it necessary to include at least three shells of \textbf{G} vectors in the expansion to correctly reproduce $V_p$ obtained from first-principles calculations. The constructed $V_p$ and the $ab$ $initio$ $V_p$ are shown in Figs. \ref{fig:si_fig3}(a) and (c) for $\theta=0.82\si\degree$ and $\theta=0.33\si\degree$ respectively, depicting good agreements.

The out-of-plane dependence of the $V_p$ is described by \cite{zhao2021universal}:
\begin{align}
    V_P(\textbf{R},z) = \textrm{sgn}(z)\sum_{\textbf{G}}V_{\textbf{G}}e^{i\textbf{G}.\textbf{R}}e^{-G|z|}
\end{align}
where $G$ is the magnitude of $\textbf{G}$. The variation of the $V_P(\textbf{R},z)$ is shown for $\theta=0.82\si\degree$ and $\theta=0.33\si\degree$ in Figs. \ref{fig:si_fig3}(a) and (c) respectively. As expected the potential decay becomes slower with increasing domain size. We also show the variation of the out-of-plane electric field ($E_z=\frac{\partial V}{\partial z}$) with increasing $z$ for those two twist angles (Figs. \ref{fig:si_fig3}(b) and (d)). Since we observe that the DW width does not change significantly with $\theta$ for  $\theta \lesssim 1\si\degree$, the electric field magnitude is similar at DW when $z$ is very close to the interface and decays with increasing $z$. However, the decay of the electric field is slower than that of the corresponding potential. 

\bibliography{ref25}